\documentclass[aps, pra, reprint, noshowpacs, superscriptaddress,nofootinbib,longbibliography,floatfix]{revtex4-2}
\usepackage{graphicx}
\usepackage{array}
\usepackage{url}
\usepackage{multirow}
\usepackage{soul}
\usepackage{xcolor}

\usepackage{mathtools,amssymb}
\usepackage{tikz}
\usepackage{tabularx}
\usepackage{enumitem}
\usepackage{censor}
\usepackage{svg}
\usepackage{longtable}

\newcolumntype{L}{>{\raggedright\arraybackslash}X}
\newcolumntype{C}{>{\centering\arraybackslash}X}

\newenvironment{iquote}
    {\vspace{-.5\baselineskip}\itshape\list{}{\leftmargin=0.2in\rightmargin=0.2in}%
    \item\relax}
    {\endlist\vspace{-.5\baselineskip}}
\newcommand{\myquote}[1]{\vspace{-.1\baselineskip}\begin{iquote} 
#1\end{iquote}\vspace{-.1\baselineskip}}

\definecolor{myred}{rgb}{.85,0,0}

\begin{document}
\title{Development of a global landscape of undergraduate physics laboratory courses}

\author{Gayle Geschwind}
\affiliation{JILA, National Institute of Standards and Technology and the University of Colorado, Boulder, CO 80309, USA}
\affiliation{Department of Physics, University of Colorado, 390 UCB, Boulder, CO 80309, USA}

\author{Micol Alemani}
\affiliation{University of Potsdam, Institute of Physics and Astronomy, Karl-Liebknecht-Str. 24-25, 14476 Potsdam, Germany}

\author{Michael F.J. Fox}
\affiliation{Department of Physics, Imperial College London, Prince Consort Road, London, SW7 2AZ, UK}

\author{P.S.W.M. Logman}
\affiliation{Leiden Institute of Physics, Leiden University, The Netherlands}
\affiliation{Leiden University Graduate School of Teaching, Leiden University, The Netherlands}

\author{Eugenio Tufino}
\affiliation{Department of Physics, University of Trento, Italy }

\author{H.J. Lewandowski}
\affiliation{JILA, National Institute of Standards and Technology and the University of Colorado, Boulder, CO 80309, USA}
\affiliation{Department of Physics, University of Colorado, 390 UCB, Boulder, CO 80309, USA}

\begin{abstract}
Physics Education Research (PER) is a global endeavour, with a wealth of work performed at a variety of institutions worldwide. However, results from research into undergraduate physics laboratory courses is often difficult to compare due to the broad variations in courses. We have created a survey to help classify these courses to compare and contrast them, which will be useful in two key endeavours: comparisons between PER studies and providing useful data for individual instructors hoping to improve their courses, thus providing information relevant to both researchers and instructors.  While we are still in the process of collecting sufficient data to create a full taxonomy of laboratory courses, we present here details of the survey creation itself, as well as a first look at the data collected, which includes a broad landscape of lab courses in 41 countries. We used both quantitative and qualitative methods in analyzing the data collected. Some of these results include similarities between courses, such as students often using pre-constructed apparatuses and that instructors hope for students to learn technical skills. We also find differences in courses, such as in the number and types of goals of the course, as well as the activities students participate in.
\end{abstract}

\maketitle

\section{Introduction}
Physics education is a global endeavour. As we strive to find the best methods of education for the next generation of physics students, undergraduate physics education can benefit from an international perspective, both for improving and comparing courses and to aid students studying worldwide. 

In today's world, international collaboration is growing due to more accessible technology and the need to engage scientists across the world to answer important questions and solve critical issues that are global in nature~\cite{Monastersky2019}.  Thus, physics education should cross the boundaries of countries to form a cohesive structure to enhance education of future scientists. In order to best conduct physics education research to improve education, we first need to understand the similarities and differences in how physics is taught worldwide, including degree requirements, classroom environments, and experiences of students. This will also help future collaborations amongst physicists, as they can better understand their collaborators' previous educational experiences, which could lead to a better appreciation of the variety of backgrounds of participants in a collaboration.

Here, we focus on the context of undergraduate physics laboratory courses due to the significant current work in this space~\cite{Caballero2018,LabsFocus,Hofstein2004,May2023}, as well as the importance of physics laboratory courses in general and the unique skills they can provide for students~\cite{Kozminski2014}. One important step in the process to improve physics laboratory education is to  understand what these courses currently look like globally.

Towards this end, our ultimate goal is to create a taxonomy, or classification scheme, of undergraduate physics laboratory courses that can be applied worldwide. This taxonomy would have numerous applications, including to gather information about courses so that lab instructors and course developers may be inspired by others, as well as to facilitate comparisons that may be made through Physics Education Research (PER) studies. From a research perspective, it is difficult to know whether studies done in certain courses can be compared to others, as lab courses are a rich and complex space with a large variety of implementations. A taxonomy could help classify these courses, so research results can be used appropriately without over-generalization. 

A taxonomy could also be used to standardize comparison data that is currently presented in reports to instructors about their courses from research-based assessment instruments (RBAIs)~\cite{Wilcox2016_5,VanDusen2021}, assessments used to determine how well students collectively are meeting course learning goals (as opposed to assessments used to individually evaluate students)~\cite{Madsen2017}.
Typically, results from RBAIs for laboratory courses present instructors with both their own students' performance, as well as data from other courses for comparison. Unfortunately, these comparison data usually include data from all students who have taken the assessment, regardless of the type of course it was used in. This makes it difficult to know whether the comparison data is appropriate. A taxonomy could be used to select comparison data from only similarly characterized classes. 

On the practical side, a taxonomy could allow instructors to learn from one another about what they do in order to improve or transform their courses. One could also use a taxonomy of lab courses to learn about student experiences in different systems; this may be helpful in graduate admissions, as students apply to study across the globe coming from undergraduate institutions around the world. Further, a taxonomy scheme could help facilitate international collaborations amongst physicists with different educational backgrounds.

The specific goals of this paper are as follows:
\begin{enumerate}

\item Provide initial validation related to the development and validation of a survey designed to capture information about undergraduate physics lab courses worldwide

\item Report initial results from the survey to demonstrate the diversity of lab courses that the survey is able to capture

\end{enumerate}

In terms of our first goal, we developed a survey that we validated with interviews and then distributed to obtain responses to analyze to get a view of lab courses. Here, we present a detailed view of the development of the survey, as well as information related to validity. Specifically, we discuss in detail content validity, construct validity, and face validity of the survey. To achieve the second goal, we present a first look at the similarities and differences of 217 lab courses in 41 countries represented in the initial data collections. 

The end goal of this work is create a taxonomy of lab courses, which will be possible with additional data collection. The project goals also include gathering input through interviews with lab instructors across the world to both understand the scope of lab instruction, to gather input to the development of the research-based survey, which aims to capture the structure, goals, and activities in lab classes throughout the world, and to make sure the survey is interpretable for both native and non-native English speakers.

Future efforts will work to collect significantly more survey responses to be able to apply clustering methods~\cite{Everitt2011} to create a taxonomy of labs courses that can be used by both instructors and education researchers. 

\section{Background}

\subsection{Prior PER on Laboratory Courses with a Global Perspective}

PER in undergraduate physics laboratory courses is less common than research investigating lecture courses, though research focused on laboratory courses is growing in popularity~\cite{LabsFocus}. Though undergraduate laboratory courses are often the only time students might get experience with experimental physics, these courses are frequently overlooked as less important than their lecture counterparts in the curriculum overall~\cite{Etkina2023}, as often students complete fewer lab courses than lecture courses. Further, laboratory instruction can be challenging due to personnel limitations, financial constraints, aging equipment, outdated experiments, and lack of advanced courses~\cite{Feder2017}. Despite these challenges, students have the opportunity to learn valuable skills in laboratory courses --- such as hands-on technical skills, experimental design, modeling, troubleshooting,  and data analysis techniques --- that are often not otherwise covered in the physics curriculum~\cite{Kozminski2014}. Because these courses can be critical to the development of students' experimental skills, knowledge, and habits of mind and are often resource-intensive, it is vital to ensure that the courses are meeting their goals with the support of research-based practices and assessments being developed by researchers in PER~\cite{Feder2017}. 

Currently, the field of PER in laboratory courses is focused largely in the United States, but is quickly becoming more international. In this section, we highlight some contributions to the PER literature from the community outside of the USA. For example, one paper from Taiwan discusses methods of integrating technology into physics lab courses, including potential options for virtual and remote laboratory instruction, including the creation of a framework for others hoping to use technology to support inquiry-based activities~\cite{Chen2012}. A recent paper from Finland discusses modernization of a physics lab course at the University of Helsinki, with details about shifting to more open-inquiry activities~\cite{Kontro2018}. Other work from Finland details methods of assessing students work in lab courses, including different types of examination and feedback~\cite{Ketonen2023}. Recent work from India discusses shifting an undergraduate electronics lab towards open-ended activities in order to help improve students' research skills, including technical skills, problem-solving abilities, and collaboration~\cite{Narayanan2023}. One paper from Germany presents a similar shift from prescriptive laboratory activities towards a skills-based course with more authentic experiments, finding that students are more engaged and are better able to master important laboratory skills, such as keeping a lab notebook~\cite{Alemani2023}. Further, in Italy, a group of researchers examined the transformation of a lab course to include activities with Arduinos and smartphones, including an open-ended aspect~\cite{Organtini2022}. Other work, from the Netherlands, shows a tendency towards open-inquiry lab courses~\cite{Logman2021, Pols2023, Logman2024}.

In addition to these efforts in in-person labs, the COVID-19 pandemic prompted research into remote laboratory activities. One such paper from the Netherlands described this abrupt transition where they investigated the use of Arduinos with open-inquiry activities~\cite{Bradbury2020}. An additional paper from France also details the use of Arduinos in a project-based lab course for third year students, in which they are given complete decision-making control over their experimental setup and what to investigate with the Arduinos in order to help them learn more about the nature of experimental physics~\cite{Bouquet2017}. Further work from the Netherlands reports the creation of a Mach-Zehnder interferometer from children's toys with an Arduino detector. This work highlights the ability to achieve experimental physics learning goals without the need for expensive resources~\cite{Feenstra2021}.

In addition to research from individual countries, collaborations between researchers in different countries have become more common as modern technology allows us to connect with people around the world. For example, a collaboration between universities in Finland and the United States investigated students' abilities in critical thinking over the course of a semester~\cite{Pirinen2023}. Another collaboration involving researchers in Germany, Finland, Switzerland, and Croatia examined the use of digital experiments in physics lab courses, including development of a questionnaire in four languages to investigate student use of these online experiments aimed at remote learning~\cite{Lahme2023}. Additionally, researchers in Germany and the United States investigated student views of experimental physics in German lab courses, finding distinct differences between these students and their American counterparts.~\cite{Teichmann2022}. Another collaboration between researchers in Denmark, Czechia, and Slovenia explored teacher education regarding lab courses with global input from a discussion at a conference. This work specifically focused on learning goals and the role of labs in teaching physics~\cite{Bearden2022}.

These collaborations and international studies can reveal similarities and differences in the ways that lab courses are taught around the world~\cite{Pols2021}, allowing us to challenge our perspectives on how lab courses are taught. However, it is difficult to directly compare research from all of these studies without an understanding of the basic laboratory course structure, goals, and activities at these different institutions. Even within the United States, laboratory courses differ vastly between different colleges and universities~\cite{Holmes2020}; adding an international component to the mix further complicates this. Some prior research has investigated physics lab instruction in North America~\cite{Holmes2020} using course instructor surveys from RBAIs, but this does not include a broader international aspect. Holmes et. al found that physics lab instruction across North America varies considerably in many aspects, including course goals, activities, and pedagogical methods~\cite{Holmes2020}. Another recent paper compared instructional strategies during the pandemic from one university from each of the three countries (United States, Sweden, and Australia) and found that, despite the widely varying locations and cultural constructs, all universities struggled with successfully implementing emergency remote instruction~\cite{Salehi2023}. 

One class of studies that aims to be broader than just including a few countries is the use of RBAIs. These assessments are intended for gathering information about students' performance collectively, rather than for assigning individual students grades. They are frequently used to determine whether a course is meeting its learning goals~\cite{Madsen2017}. One such RBAI is the Colorado Learning Attitudes about Science Survey for Experimental Physics (E-CLASS)~\cite{Zwickl2013b, Zwickl2014,Wilcox2016_2,Wilcox2017}. Despite the name, this assessment instrument has been used broadly with greater than  100,000 total responses from many countries (mostly in North America, Europe, and Asia). It has also been translated to several different languages (including Swedish~\cite{Henriksson2020}, Italian~\cite{Organtini2022}, German~\cite{Teichmann2022}, Chinese, Spanish, Hebrew~\cite{Levy2020}, Norwegian, and Amharic) to allow for easier administration in other countries. Another globally used RBAI, the Physics Lab Inventory of Critical Thinking (PLIC)~\cite{Walsh2019,Walsh2018,Quinn2018,Holmes2015_2}, which focuses on critical thinking in physics labs, including questions related to data analysis and measurement uncertainty. This survey is available in Chinese, Finnish~\cite{Pirinen2023}, German, and Spanish, in addition to English. However, even if the data collection using these RBAIs includes many countries, they did not include input from more than a few at most in the development of the instruments. 

\subsection{Prior work on characterizing undergraduate physics courses}

There has been some previous work on characterizing undergraduate physics courses that we can build on for the current study.

Prior research has worked to characterize physics theory courses, especially at the introductory level. For example, some research focuses on introductory physics problems and creation of a taxonomy of the cognitive processes required to solve these problems~\cite{Teodorescu2008, Teodorescu2013}. This taxonomy can be applied to help develop assessments in courses, guide curriculum development, and determine learning goals for introductory courses. However, little research has been done to apply such a taxonomy to higher-level theory courses or to compare theory courses between different institutions.

Some recent research has focused on lab courses. One effort to characterize laboratory courses is related to the administration of E-CLASS and PLIC. To implement E-CLASS using the centrally administered version in English, instructors complete a course information survey, which gathers information about the course itself, including information about the level (introductory or beyond introductory), whether it is algebra- or calculus-based, the number of students and staff, and how frequently students participate in various activities. A nearly identical  set of questions is used in the administration of the English version of PLIC~\cite{Holmes2020}.  While this information is useful in many research applications, it is limited in scope and does not provide enough for course classification  for worldwide applications~\cite{Wilcox2016_4, Holmes2020}. For example, it is missing information about activities, pedagogies, and course design. Some questions from the course information surveys for E-CLASS and the PLIC were used as input for the survey discussed in this paper, including questions about the level of the course (introductory versus beyond introductory), type of institution, and information about course goals~\cite{Holmes2020}, though these current instruments lack important details about courses, hence the need to develop a new instrument.

Another effort we draw from was initiated during the COVID-19 pandemic, where research was conducted regarding the switch to emergency remote laboratory instruction. Researchers created a survey to collect information from course instructors about the structure, goals, and components, as well as other features of their remote or hybrid lab courses~\cite{Borish2022, Fox2021, Werth2022}. The pandemic instructor survey included some similar questions to those in the E-CLASS course information survey, but also included questions about changes to course goals and activities (whether they were incorporated in the course both before the pandemic and while the course was remote). While the data were helpful in analyzing differences between instruction pre-pandemic versus during the pandemic, the questions were not extensive enough to fully characterize laboratory courses and were created locally on an extremely short timescale to capture the rapidly evolving situation.

\section{Methods}
This section discusses the various methods used in the paper. First, we discuss the authors' backgrounds as relevant to this paper. Next, we present information about the survey's creation, including the initial brainstorming steps and organization into a logical survey format, followed by interviews with instructors for validation of the survey. Finally, we discuss the dissemination of the survey broadly, including the limitations of our data collection.

\subsection{Author Professional Positionality Statements}

As the work presented here relies, in part, on the professional experience and networks of the authors, we present our backgrounds relevant to the work. Further, the backgrounds of the authors are relevant in our first steps towards validation.

Author G.~G. is a Ph.D. student currently working in PER on the development, validation, and analysis of a measurement uncertainty focused RBAI for use in undergraduate physics laboratory courses. She has prior experience both as an undergraduate and as a graduate student working in experimental physics; her undergraduate career focused mainly on experimental atomic, molecular, and optical physics while her graduate experimental physics experience is in biophysics. Finally, she has been a teaching assistant in a sophomore-level undergraduate physics lab course at the University of Colorado Boulder. She was not present at the workshop where the initial idea for the taxonomy work originated, but was brought in some time later to work on the project.

Author M.~A. is the laboratory course coordinator at the University of Potsdam, which she completely redesigned using research based results. Her physics background is in solid state physics, but she is now focusing on PER. During her PhD she built a low temperature scanning tunnelling microscope and used it to investigate the manipulation of isolated molecules on surfaces. During her postdoc, she studied electrical transport properties of graphene. She worked as an experimental physicist both in US and Germany and taught lab courses in both these countries. As a student she did her lab courses in Italy. She is an active member of the working group about undergraduate laboratory courses called Arbeitsgruppe Physikalische Praktika (AGPP)~\cite{AGPP} of the German Physical Society (DPG) and since 2021 serves as board member for the physics education section of the DPG. 

Author M.~F.~J.~F. has been the second-year laboratory course coordinator at Imperial College London since 2023. His research is focused on student learning in undergraduate teaching laboratories and equity in physics. He completed a PhD in theoretical plasma physics, taught high-school physics in London for 3 years, and worked as a postdoctoral researcher with H.~J.~L. in PER related to laboratory courses completing work on the quantum industry~\cite{Fox2020quantum}, as well as analysis of E-CLASS data~\cite{Fox2021} and development of the MAPLE survey~\cite{Fox2020}. 

Author P.~S.~W.~M.~L. is the laboratory course coordinator at the Leiden Institute of Physics in The Netherlands. His physics background is in applied physics, but since 2009 he is focusing on physics education. From 1993-2014 he worked as a high school physics teacher. In 2014 he finished his PhD in physics education in which he used educational design research to develop a teaching-learning sequence on the general law of energy conservation in which students reinvent that law by performing various experiments. After finalizing his PhD he started work in Leiden where he has been redesgining the undergraduate lab courses since 2017 using educational design research. He is a board member of the Groupe International de Recherche sur l'Enseignement de la Physique (GIREP - International Research Group on Physics Teaching)~\cite{GIREP} and co-leader of the GIREP Thematic Group on Laboratory Based Teaching in Physics (LabTiP)~\cite{GIREP_Group}.

Author E.~T. is a Ph.D. student actively involved in PER. His research focuses on the integration of active learning methods in both introductory university physics courses and high school settings (in particular, using the ISLE approach). With many years of experience as a high school physics teacher, he has developed a keen interest in the use of digital technologies to enhance physics learning. He has been involved in the introduction of the E-CLASS assessment tool in Italy. E.T. attended the workshop in April 2022, where he presented the first results of the implementation of E-CLASS in Italy.

Author H.~J.~L. is a physics professor who runs two research groups, one in PER and one in experimental chemical physics. Her PhD was in the field of experimental atomic physics (Bose-Einstein Condensation), and her postdoc was in experimental molecular physics (Cold molecule spectroscopy). She has over 25 years experience designing, constructing, and using table-top experimental apparatus. Her work in PER began in 2010 and has focused mostly on laboratory courses at the undergraduate level. Besides the current work, she has had many international collaborations in PER, including with researchers from China, Oman, South Africa, Denmark, Norway, United Kingdom, Italy, and Germany. Additionally, she has taught all of the undergraduate physics laboratory courses at the University of Colorado numerous times. She also led efforts to transform three of these courses through research-based practices. She has served on the Board of Directors of the Advanced Laboratory Physics Association (ALPhA)~\cite{ALPHA} for 11 years, including two years as President of the organization.  She was present at the workshop in April 2022 that initiated this project.

\subsection{Survey Creation}

\subsubsection{Initial Creation of Ideas and Organization}
The project idea emerged from a small workshop (about 20 attendees) at Imperial College London in April 2022, where PER researchers who had used the E-CLASS survey in Europe and Western Asia were invited to present. From discussions amongst attendees, it became apparent there was not a clearly defined way to compare lab courses from different countries. A subset of attendees decided to embark upon a project to create a lab taxonomy that would be broadly applicable. The idea would be to create a valid survey to collect data on lab courses around the world and use those data to create a taxonomy.

The first step towards development of the survey was a collective brainstorming session over Zoom, where we discussed the most important aspects of undergraduate physics laboratory courses and created a virtual whiteboard to collect and partially organize the ideas. Many of the ideas organized on the whiteboard, which eventually morphed into the various categories in the survey, were based on the Spinnenweb (spiderweb) representation of curriculum and learning proposed by van den Akker~\cite{VanDenAkker2003}. This model examines why students learn in different facets. A large fraction of the survey was modeled off this structure, including sections such as grouping, assessment, goals, activities, content, instructional staff roles, and materials and resources.

After this initial process, the results of the whiteboard activity were loosely organized into a document by category. This document included many of the ideas that eventually went into the survey, but was missing several important concepts and  included many items that we decided to remove. For example, a list of equipment that students might have access to in the course was removed for being too unwieldy for both survey participants and researchers to handle. We also added questions from previous E-CLASS/PLIC and pandemic instructor surveys.

This list of lab course components, along with feedback provided by the authors, based on personal experience and the literature, and external physics education researchers, was transformed into a survey format. This involved organizing the information we wished to probe into meaningful categories,  ordering those categories in an intuitive way, and turning ideas we wanted to probe into questions. All authors iterated on the survey several times until a cohesive final product emerged. Many of the iterations involved language changes, with interviews (as detailed below) later validating the wording choices made for the survey. At this stage, the survey document was coded into a Qualtrics survey in order to prepare for the interview validation phase of survey development. 

\subsubsection{Interviews with Lab Instructors}

In order to validate the survey, we performed interviews with lab instructors. These interviews provided evidence for three types of validity: construct validity, content validity, and face validity. Construct validity aims to determine if the survey measures the concept it is intended to; in our application, we analyze whether the survey addresses information about undergraduate physics lab courses. Content validity aims to determine if the survey is fully representative of what it aims to measure, which in this case means that we would like to determine if the survey fully spans the space of relevant information about lab courses. Finally, face validity determines whether the content of the survey is suitable for its goals --- that is, if the questions in the survey appear to measure information regarding these courses. 

Interviews were solicited from contacts known by the authors. These contacts were compiled into a list and solicitations were chosen such that only one person per country would be solicited with a reasonable worldwide spread of country participants. We chose to exclude Germany, Italy, the Netherlands, and the United Kingdom from interviews due to already having researchers on our team from these countries. We did include the United States in interviews due to the diversity in universities in this country~\cite{McCormick2020, Usher2010, Altbach2010, Ziegele2013, McCormick2005}. We conducted interviews only with those who currently teach or have previously taught undergraduate physics laboratory courses. 

In total, we conducted 23 interviews with participants from 22 countries; the United States was sampled twice. A map of the countries participating in interviews is shown in Fig.~\ref{fig:interviewmap} and a list of these countries can be found in the Appendix. We solicited 32 participants from 26 countries in total, leading to approximately a 72\% response rate. Interviews were solicited in several rounds, and we determined after 23 interviews that changes to the survey had become minimal and therefore, further interviews would be unnecessary. (We made changes to the survey after nearly every interview and would present the newest version to the next interviewee).

Author G.~G. conducted all of the interviews over Zoom. All interviews were conducted in English. Each interview lasted between 33~\textendash~76 minutes. Both video and audio of the interviews were recorded for later analysis. Additionally, the interviewer took notes during the interview and, in most cases, these notes provided the basis for further changes to the survey. Interviewees were provided with a link to the survey during the interview and were asked to talk through the questions while sharing their screen. They were instructed to consider all undergraduate physics laboratory courses they had knowledge of when considering whether the questions made sense, as well as whether the questions spanned the space of all knowledge they felt was important  to collect. 

The interview protocol was created to validate the survey questions by answering three validation questions:

\begin{enumerate}

\item Are the survey questions understandable to those who are not from a country represented by an author? Are the survey questions interpreted in the way they were intended?

\item Do the survey questions make sense in the context of courses the instructor has taught?

\item Do the survey questions fully span the space of information that the instructor feels is important to capture about undergraduate physics laboratories?

\end{enumerate}
For most survey questions, interviewees were asked whether they understand the question, as well as whether anything was missing; in some cases, interviewees were also asked to explain in their own words the concepts and ideas presented in certain questions, especially in the goals and activities sections of the survey.

In terms of the first validation question, we probe whether the (mostly) American English used throughout the survey is understandable both to those who have learned English as a second language, as well as to those who speak English as their primary language, but know a different dialect (British, Canadian, or Australian English, for example). Differences arise due to variations in terms used by British English (e.g.,`revise' means `study' in British English, but means `alter' in American English), as well as terms not commonly used for those who speak another language as their primary language (e.g., the term `rubric' was determined to be unfamiliar to many non-native English speakers, but when the concept was described, interviewees were familiar with it). Because the entire survey is in English, the interviews were also conducted in English. While this limits the pool of both interviewees and those who can participate in the survey, and certainly places limits on some countries more than others, translating the survey is currently outside of the scope of this project.

To answer the second of the validation questions, interviewees were asked about whether each survey question was applicable in courses familiar to them, thus providing evidence of construct and face validity. In some cases, questions had to be added or logic had to be introduced in order to ensure that people taking the survey would be able to appropriately answer all questions without confusion due to certain questions not being applicable to their courses. For example, in one case, we added a question about whether the lab course meets weekly. In the case that it does, the instructor is sent to a page with the original questions about number of hours per week the course meets and the number of weeks the course runs. In the case that the course does not meet weekly, instructors are sent to a different set of questions in which they are asked to describe their course meeting (how often, how many hours per term, etc.). 

Answering the third validation question helped us to add, remove, and revise questions to ensure we gathered all of the information we need to create a taxonomy and describe the state of undergraduate physics laboratory courses around the world, thus providing evidence of content validity. Interviewees were asked whether questions regarding goals of the course and activities students participate in were complete or missing important relevant information. Many times, interviewees suggested additions that helped to make the survey more accurately capture the full breadth of components of the courses. In addition to ensuring each individual question fully spanned the space of information we hoped to gather about that specific topic being probed, interviewees were asked at the end of the survey whether they felt anything was missing that they thought should have been included. 

Many changes to the survey occurred concurrent with the interview process. As interviewees made comments about the survey, changes were discussed with members of the team and implemented so that updated questions could be validated during the interview phase. More details of the results of this process are discussed in Sec. \ref{sec:Lab Tax Survey}.

\begin{figure*}[ht]
    \centering
    \includegraphics[width=\textwidth]{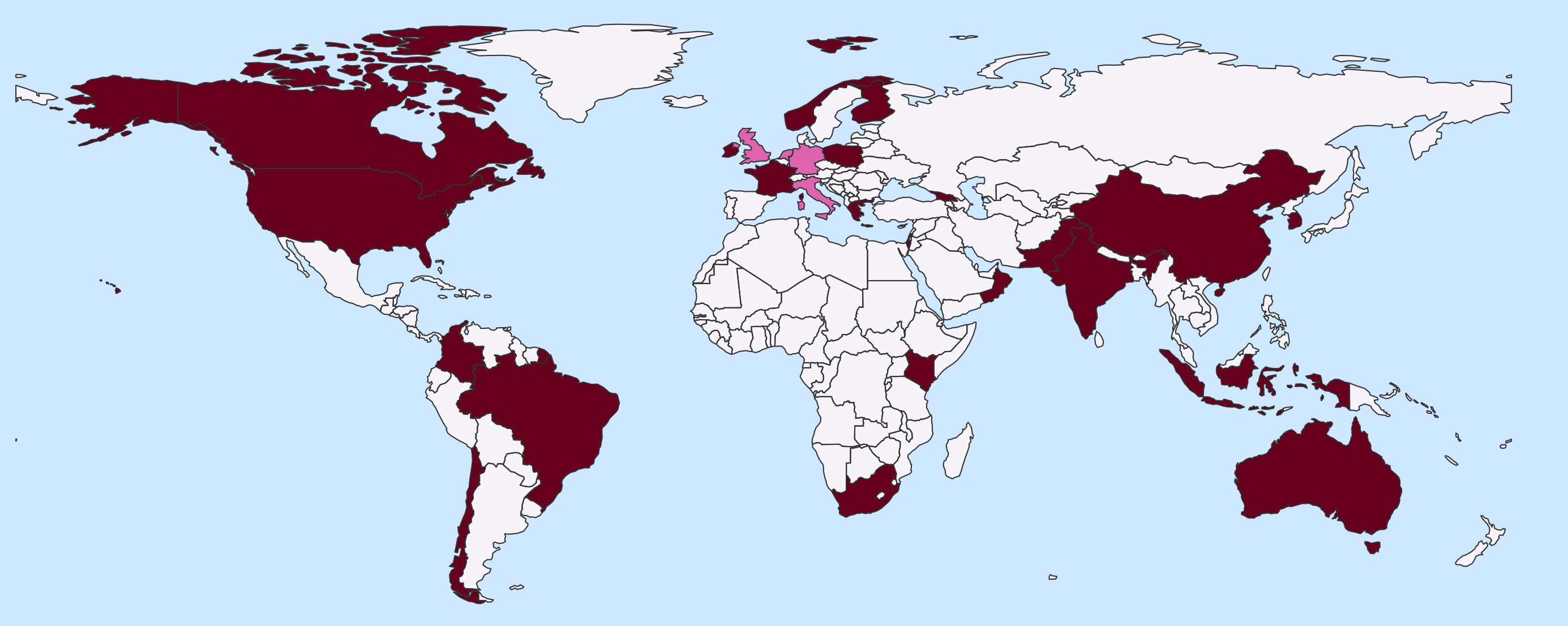}
    \caption{World map indicating the countries where we interviewed lab instructors. Each maroon country had one instructor interviewed, aside from the United States, which had two. The pink countries represent the home countries of the authors of this paper, aside from the United States, and therefore no interviews were solicited from instructors in these countries because the authors could provide the necessary feedback. We interviewed 23 instructors from 22 countries. A full list of these countries can be found in the Appendix.}
    \label{fig:interviewmap}
\end{figure*}

\subsection{Survey Dissemination, Data Collection, and Data Analysis}

We first disseminated the final version of the survey using all of the international contacts of the authors. We compiled a list of our contacts ($\sim$130), then emailed a solicitation to take the survey to people we knew through our professional networks. This solicitation invited people to take the survey, as well as to pass it along to others they know (whether in their own department or at other institutions), which is often referred to as snowball sampling~\cite{Goodman1961, Biernacki1981}. We included snowball sampling as it is typically used for difficult-to-access populations. In this case, physics lab instructors are such a population as their positions and titles vary and contact information is often not readily available, as previously discussed in another context~\cite{Dana2023}.

We also posted the survey on a variety of listservs relevant to lab instructors. These included the ALPhA listserv~\cite{ALPHA}, two American Physical Society (APS) discussion boards (Forum on Education~\cite{APS_FED} and Topical Group on Physics Education Research~\cite{APS_GPER}), a newsletter distributed to members of GIREP, and a JISCmail forum for physics education researchers and instructors in the United Kingdom~\cite{JISCmail}. Further, the survey was advertised during the ALPhA Beyond First Year (BFY) IV Conference, American Association of Physics Teachers Summer Conference, Physics Education Research Conference, and at the GIREP Conference (all in July 2023) during various authors' talks and poster presentations.

Next, we compiled a list of 171 countries worldwide and focused our efforts on those not represented in our sample thus far, especially those within world regions not well-represented. Each author then searched for publicly available information about institutions with physics departments (e.g., from institution websites) within 10~\textendash~15 unique countries and contributed to a database with contact information for department heads and lab instructors. This approach has limitations, as many institutions did not have contact information easily accessible. However, this led to a large increase in responses: we received about 53 additional responses from 24 countries to add to our data set. The solicitation e-mail we sent to these institutions requested that they send to their contacts as well. In selecting countries, we chose places where we did not have a well-represented sample, such as much of Asia, Eastern Europe, South America, and Africa. This helped spread the survey to a more global audience than just where our contacts were located.

Data presented here were collected from 20 June 2023 until 10 January 2024, though the survey is still open and collecting responses~\cite{LabTaxSurv}. We include data only from people who responded to a minimum number of questions (about 80\% of the survey). This removes 121 responses with no questions answered and 18 responses that answered only some questions, leaving 217 out of the initial 356 responses. Because we do not force responses to most questions on the survey, aside from a couple that are necessary for future logic within the survey, each question has a varying number of responses. The number of respondents is reported for each part of the results section as appropriate. The median time to take the survey was 18 minutes.\footnote{Median is reported to exclude the outliers of those who leave the survey open for multiple days without actively filling it out before submitting it, thereby skewing the mean and making it an inappropriate statistic to report.} We present descriptive statistics of the survey responses in order to answer our second research question, as well as to provide evidence that our survey is able to capture a wide variety of courses from multiple countries. 

Finally, in analyzing the data, we compared responses to two questions, one regarding the goals of the course and another regarding the activities students participated in during the course in order to determine if these activities align with the course goals. Similarly, we also matched these course goals with items that might be used to evaluate students for their final course grades. Authors G.~G. and H.~J.~L. along with an outside PER postdoctoral researcher worked together to match the course goals with activities, as well as the course goals with items graded. We then determined whether instructors are connecting their course goals with the activities students participate in and in the ways they are evaluated in the course. As a simple example, the course goal ``Developing lab notebook keeping skills'' can be matched with activities ``maintain an individual lab notebook'' and ``maintain a group lab notebook''. This goal can be matched with ``Lab notebooks'' under items graded. After G.~G., H.~J.~L., and the outside researcher came to agreement about matching, an analysis was done to determine how many activities instructors chose that matched the course goals, as well as how many items graded.

\subsubsection{Limitations of our Data}

First, aside from the United States, the countries where the authors are located (Germany, Italy, the United Kingdom, and the Netherlands) are oversampled based on the number of institutions present in each country, while most other countries are undersampled by this same metric. For example, we have only three responses from China and two responses from India, two largely populated countries with strong physics programs. We also have only four responses from Africa and nine responses from South America, therefore undersampling large areas of the world. We are also missing many countries entirely. The United States is underrepresented compared to responses received from other countries based on the number of higher education institutions --- we have only 63 responses for the USA. Additionally, within the United States, there are a large variety of types of institutions~\cite{McCormick2020, Usher2010, Altbach2010, Ziegele2013, McCormick2005}, so getting a truly representative sample of institutions would be difficult and is something we do not currently have at this point in our data collection.

Second, some of our responses are clustered at specific institutions. One example of this is Canada --- while we have six responses from Canada, five of them are from one university within Canada and all six responses are from a single province.

Third, since the survey was disseminated primarily using our contacts and listservs we are members of, those who responded are more likely to be interested or involved in physics education research to some extent. This also biased the countries from which we received responses, as those with relationships with the authors were more likely to fill out the survey and pass it on to their colleagues, hence why the country bias is skewed towards our home countries. There is also bias in who chose to fill out the survey: those with interest in improving their programs and are invested in the quality of lab teaching are more likely to fill out the survey. 

Fourth, the survey is available only in English. While many of our colleagues do speak English, we miss many people who do not know English well enough to complete the survey. We chose not to translate the survey at this time due to constraints on resources and expertise.

Finally, because of these limitations, we do not present uncertainties in our tables in most cases. Due to the sampling bias in our data, our unknown systematic error is likely larger than the uncertainty determined by statistical means. It could be therefore misleading to present the statistical uncertainty.

Considering all of the limitations of our dataset, we advise caution when interpreting results. While the development of the survey and validating it are not subject to these limitations, the data addressing our second goal of this paper --- providing an overview of undergraduate physics lab courses globally --- is affected by them. In particular, strong trends are likely to be applicable, but weaker trends and subtle differences shown in the data may not accurately represent the full global landscape of labs.

\section{Results and Discussion}
\label{sec:results}

\subsection{Lab Taxonomy Survey and Validation}
\label{sec:Lab Tax Survey}

The validation of the final version of the survey, as well as the survey itself are major results of this work. The survey was developed through a systematic process and extensive interviews, and will hopefully serve as a foundational tool for years to come. An adaptation of the Qualtrics version of the survey is presented in the Supplementary Materials.

As described previously, we have evidence of validity of the survey along three axes: face, construct, and content validity. Face validity is mainly an aspect that determines whether the questions appear related to the survey's goals. Face validity was determined by the authors during the initial preparation of the survey questions, while further evidence was provided during interviews in that interviewees noted that the questions generally made sense in context and all were related to the goal of gathering information about lab courses. None of the questions were seen as outside of the scope of the survey by any of the interviewees.

Next, construct validity was also determined during  survey development. The ultimate goal of the project is the creation of a taxonomy of courses; however, we do not have enough data to create this scheme yet and, therefore, do not have all of the evidence required for full construct validity. However, the interviews did provide evidence that this type of validity is present in our survey, as the questions measured information about their courses according to the interviewees. 

Finally, we determine content validity primarily through the interviews. Here, the goal of determining whether any questions are missing from the survey was our main method of providing content validity. When interviewees did respond that things were missing from the survey, edits were made to include these items and therefore ensure the survey is probing all important concepts related to information about lab courses.

The survey is delivered online via Qualtrics and remains open for data collection~\cite{LabTaxSurv}. We note that the survey does not collect information about the instructors (e.g., name, e-mail address, demographics), but rather focuses only on the course itself. 

The survey is structured with eight separate sections to capture a variety of information including: overall course and institution characteristics, students, grouping of students, instructional staff, goals, activities, evaluation, and an optional open text box for any additional items, including a suggestion for lab activity titles.

Interviews helped shape the final form of the survey. The first general category of survey edits were simple improvements, including the addition of a progress bar, bolding `select all that apply' wherever it appeared in order to draw attention to it (based on several interviewees missing this text in the questions), inclusion of a back button, and other general readability improvements. These were minimal edits that did not significantly change the contents of the survey, but rather improved the user experience. In addition to these changes, minor wording changes and clarifications were made throughout the survey to improve understandability to a wider audience. We also note that for each section, the number of questions did not change significantly during interviews, but rather the content and wording of the questions did.

Separately, we found that instructors might benefit from taking this survey, as well as from published works that will come from it. Some of the questions, especially those about branded approaches to instruction and the use of RBAIs, were very interesting to interviewees. These questions include links to outside resources about various instructional methods and assessments. One interviewee said:

\myquote{It looks very interesting, actually... I'm going to open all of them... that looks nice. Here, we are very far away from that... I'm going to come back to this. Thank you so much for that [sic] links,}

\noindent while examining the question about branded approaches to instruction (such as ISLE and SCALE-UP). This interviewee was excited to learn about methods they had not previously been familiar with. Questions such as these, and those about course goals and activities, can allow instructors to reflect on their courses and consider possible new teaching methods, assessments, and course goals. We hope this unintended impact of the survey can also be useful for improving laboratory instruction.

In the following, we present each section of the survey together with the description of the changes we made during their development.

\subsubsection{Overall Characteristics}

This section of the survey asks for basic information both about the institution where the course is taught and about the course. Institution questions include the location and name of the institution and highest degree it grants. Course information collected in this section includes general information, such as the name of the course, the intended level of the course (introductory vs. beyond introductory), a checklist of physics topics covered by the course, number of students enrolled in a typical term, and the basic setup of the course --- for example, whether students participate in project work, and the types of experiments students do (weekly, many experiments per week, or experiments that last longer than one week). Other questions include whether the lab course is integrated with a lecture course and whether the course includes lectures on statistics, data analysis, or experimental techniques. This section also includes questions about project work which are displayed if participants indicate that projects are part of the course. 

Many edits were made to this section during the interview phase. Several items were added to the list of topics that might be covered in a physics laboratory course, including quantum information, geophysics, and modern physics; this change occurred due to interviewees suggesting topics that they felt were important, but were not included in the original list. A question was also added to this category to probe the level of the lab, introductory or beyond introductory, after it became clear that the year of the students taking the course is not sufficient to provide this information (e.g., a course for life science majors might be mostly second or third year students, but it might be the first physics laboratory course these students take and is therefore considered to be introductory). Another question was added to probe whether the course meets weekly. If the respondents choose `no,' we added an open text box for them to provide details of their course meeting schedule. We added this text box as several interviewees mentioned that their courses followed unique course meeting schedules. Because we can not account for every such case, we determined that an open text box was the best method to collect this information and may help refine future iterations of the survey.

\subsubsection{Students}

This section of the survey first asks about students' majors and the percentage of students in the course earning a degree in physics or astrophysics. Finally, this section asks participants to estimate how many years the students have been at the university.

The most significant change that occurred in this section due to interviews was to include a question asking about the percentage of students in the course that are physics majors. We decided that simply asking about the majors of students was not enough information to determine whether the course is intended for physics majors, non-majors, or both; the inclusion of this question helps make that more clear. In addition to this change, more majors were added to the list of potential degrees students might be earning as a result of interviewee requests.

\subsubsection{Students' group work}

This section of the survey inquires about how students work in the lab: alone or with others. If students work with others, participants are asked several followup questions, including the typical size of a group, how groups are chosen, and whether students stay in the same groups for the entire course. 

No significant changes were made to this section section during the interview phase aside from slight wording edits.

\subsubsection{Instructional Staff}

This part of the survey asks about the types of instructional staff present in the lab with students. These might include faculty, lab technicians, graduate teaching assistants (TAs), and undergraduate learning assistants (LAs). In this section, participants are also asked about training provided to TAs and LAs --- both the frequency of this training and the topics covered (e.g., familiarization with equipment, pedagogy instruction, and grading training).

In this section, we added lab technicians during the interview phase at the request of interviewees. Slight wording changes were also made to the questions inquiring about the types of training.

\subsubsection{Goals}

In this section, many potential goals for a physics laboratory course are listed, and participants rank these on a Likert scale consisting of Major Goal, Minor Goal, Not a Goal, and Future Goal (not currently a goal). The development of this unique Likert scale is discussed in more detail below. 

The list of goals are as follows: 

\begin{itemize}

    \item Reinforcing physics concepts previously seen in lecture (confirming known results / seeing theory in an experiment) 
    \item Learning/discovering physics concepts not previously seen in lecture
    \item Developing technical knowledge and skills (e.g., making measurements and hands-on manipulation of equipment)
    \item Designing experiments
    \item Developing mathematical model(s) of experimental results
    \item Learning how to analyze and interpret data (e.g., linear regressions, uncertainty)
    \item Learning how to visualize data (e.g., plotting)
    \item Developing lab notebook keeping skills
    \item Developing scientific writing skills (e.g., lab reports)
    \item Developing other communication skills  (e.g., oral presentations, poster presentations)
    \item Making quick and simple approximations to predict experimental outcomes (e.g., back of the envelope calculations)
    \item Developing expert-like views about the nature of the process of doing experimental physics (e.g., experimentation is iterative, not linear)
    \item Developing collaboration and teamwork skills 
    \item Reflecting on and evaluating one’s own learning  and knowledge (metacognition)
    \item Enjoying experimental physics and/or the course 
\end{itemize}

These goals were adapted from several sources, including the AAPT lab recommendations~\cite{Kozminski2014}, the EP3 guidelines~\cite{EP3}, and previous work~\cite{Zwickl2013a}. Further, some of these goals also resulted from author brainstorming sessions as well as interviews, as described below.

The goals section went through major revisions during the interview process. First, the Likert scale was changed; initially, it included only Major Goal, Minor Goal, and Not a Goal. However, we observed that many interviewees would say that one of the goals is not a goal of their course, but they would be interested in implementing it. They would then often select `Minor Goal', despite stating it is not a goal of their course. In order to address this issue, we introduced a fourth Likert option: Future Goal (not currently a goal). While analyzing the survey data, we currently collapse this category with Not a Goal, but it helps to provide more accurate results in our data collection.

Additionally, the list of goals presented underwent revisions. Some goals, such as developing communication skills, were split into several goals [in this case, the split was into three goals: developing lab notebook keeping skills, developing scientific writing skills (e.g., lab reports), and developing other communication skills (e.g., oral presentations, poster presentations)]. This was due to interviewee input about the concepts they thought were covered by the goal. In this particular example, when asked to define ``communication skills,'' interviewees had many different ideas about what this might include. Therefore, we split this into three distinct goals in order to collect the most accurate data. Further, we added goals at the request of interviewees, such as `enjoying experimental physics and/or the course' and `making quick and simple approximations to predict experimental outcomes (e.g., back of the envelope calculations)'. 

Other changes to the goals section as a result of interviews included wording changes to clarify meaning, as well as adding examples to the goals to make them more easily understood. 

\subsubsection{Activities}

The first part of this section of the survey asks participants whether they use any officially branded approaches to lab instruction in their course [e.g., Investigative Science Learning Environment (ISLE) Physics~\cite{Etkina2007b}, Student-Centered Activities for Large Enrollment Undergraduate Programs (SCALE-UP)~\cite{Beichner2007}, and Modeling Instruction~\cite{Brewe2008}], as well as whether they use any RBAIs to evaluate the course (e.g., Survey for Physics Reasoning on Uncertainty Concepts in Experiments, or SPRUCE~\cite{Pollard2021,Vignal2023}; Modeling Assessment for Physics Laboratory Experiments, or MAPLE~\cite{Fox2020}; and E-CLASS~\cite{Wilcox2016_2}).

The next part of this section lists several activities students might participate in during an undergraduate physics laboratory course and asks how often students engage with them along the following Likert scale: Very frequently, Somewhat frequently, 1-2 times per semester/term, Would like to use in the future, and Never. Details of the development of this Likert scale are given below. The activities probed are divided into the following categories:

\begin{itemize}
    \item Data Analysis and Visualization
    \item Communication
    \item Student Decision-Making
    \item Materials and Resources
    \item Modeling and other activities
\end{itemize}

Within the above categories, examples of activities include: quantify uncertainty in a measurement, write lab reports, develop their own research questions, and calibrate measurement tools; the full list of activities can be found in the Supplemental Materials. 

The activities section also underwent significant changes during the interview process. A question probing whether research-based assessments are used to evaluate the course was added as a result of interviewees mentioning during interviews that they use some of these assessments. 

The Likert scale used in the list of activities was changed to make things clearer. Originally, the scale was: Always, Often, Sometimes, Rarely, Never. However, this scale was not appropriate for some activities. For example, students might design and present a poster once during a course, and it is unclear which of the scale points this should fall into, because it is not typical for students to make posters ``frequently'' when compared with other activities, such as quantifying uncertainty in a measurement, which might occur with every lab experiment. Similarly, it is unclear what it means to `always' complete a safety training --- some courses might require a single training, whereas others might have a few that students have to complete. 

The new Likert scale we implemented is: Very frequently, Somewhat frequently, 1-2 times per semester/term, Would like to use in the future, and Never. This scale has several advantages over the previous one. First, it includes an aspirational scale point (i.e., Would like to use in the future), which can be collapsed with Never when analyzing the survey data, but again helps discourage those who select a different option despite not using the activity (similar to the aspirational Likert scale point in the Goals section). Second, the new scale helps clarify events that might happen only one or two times in a semester, such as a poster presentation or a safety training. Finally, for activities that might happen more commonly in courses --- such as keeping a lab notebook or writing their own code --- it provides several scale points that are more easily understood. Because 1-2 times per semester/term is an option, it is clear that `somewhat frequently' means that students participate in the activity more than this, while `very frequently' indicates a higher degree. Therefore, based on interviewee responses to this scale, we feel that we have appropriate knowledge of what it means each time someone selects a particular scale point. 

In addition to a new Likert scale, many of the activities were also changed. In some cases, wording was altered or examples were added to clarify meaning. Some activities were combined into one option after it was determined that interviewees could not always tell the difference between them. One example of this is combining `refine experimental apparatus or procedure to reduce random uncertainty' and `refine system to reduce systematic uncertainty' into `refine experimental apparatus or procedure to reduce uncertainty (statistical and/or systematic)'. This was due to many interviewees not being able to give appropriate examples differentiating random and systematic uncertainty, and therefore the information obtained from probing these separately was inaccurate. 

Finally, some activities were added to this section, such as `engage with PhET simulations' and `complete safety training' due to interviewee requests.

\subsubsection{Evaluation of students' work}

This part of the survey probes how students are graded (evaluated) in the course. The first question probes whether students are assigned individual or group grades, while the second lists potential parts of the course that might factor into student grades and asks participants to select all that are used for their course. Examples include taking a quiz at home before the lab, lab notebooks, lab reports, written exams, poster presentations, practical exams (i.e., hands-on exams), and peer feedback on other students' work. Finally, participants are asked whether they use a rubric (a set of guidelines about how something is graded) to grade student work.

This section of the survey also underwent significant changes due to interviews. Initially, participants were provided with a list of items that might potentially be included in student grades and were asked to rate them on a Likert scale: Not used, Marked/graded for inclusion in final course grade, Marked/graded but not used to determine final course grade. This led to confusion, especially about certain items that are not directly used in the final grade, but might be used in some indirect way. For example, attendance at each individual course meeting might make up an overall attendance grade that is then used to determine the final course grade. Interviewees were then uncertain about where on the Likert scale to include attendance. We changed this question to be multiple response, and ask participants to select all of the items on the list that are used in grading the course. Because it is a binary option, interviewees understood how to handle indirect affects on grades (they did select these items). Removing the Likert scale helped make the survey more clear.

Additionally, some items were removed (such as reflection questions) due to interviewee lack of understanding around these points, added (such as worksheets) due to interviewee requests, and reworded to help with clarity and understanding.

\subsubsection{Optional Long Entry}

In this section, two long-form text boxes are provided. Both are listed explicitly as optional; while most of the rest of the survey's questions are optional, these are the only two questions which state this. The first asks participants to enter the titles of lab experiments in any language they wish. This is useful in looking at trends of common laboratory themes that might be present, especially in introductory labs (e.g., during the interviews, many instructors discussed using a pendulum activity in introductory mechanics). While this qualitative data might not provide the most accurate evaluation of themes --- some might choose not to include this information, and others might have laboratory titles that don't fully reflect the activities --- this will still provide a wealth of information. We encourage any language entry to allow participants to simply copy and paste lab manual titles in order to make this step easier. Online tools such as Google Translate~\cite{GoogleTranslate} provide an accurate enough translation to qualitatively code themes in later steps of the analysis. 

The second text box asks for any additional comments that participants might have. This is useful in cases where the survey may not have fully captured the experience of the participant in teaching their course or for any clarifying comments they would like to make about their prior responses. This further serves to provide content validity, as anything missing from the survey questions themselves can be captured in this box, and therefore, the survey is fully representative of what it aims to measure.

The optional long entry section was not significantly altered during the interview process.

\subsection{Survey Results}

We present here an overview of physics laboratory courses around the world based on the data we have collected thus far in order to address our second research question. We received responses from 217 unique courses in 41 countries. A figure showing a map of the countries where instructors responded to the survey is shown in Fig.~\ref{fig:worldmap}, with a full list of the number of respondents per country located in the Appendix.

\begin{figure*}[ht]
    \centering
    \includegraphics[width=\textwidth]{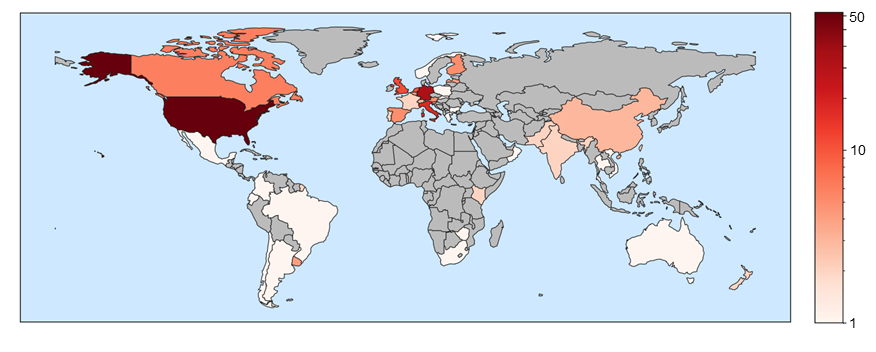}
    \caption{World Map of Number of Survey Responses. Shown on a log scale, each colored country has at least one response; countries in gray have no responses.}
    \label{fig:worldmap}
\end{figure*}

As discussed previously, our data is skewed towards the authors' countries and is lacking representation in many areas. In future work, we hope to present a more representative sample.

\subsubsection{Overall Characteristics}

Of our respondents, most courses (166/217) were offered at PhD-granting institutions, with fewer being offered at Master's-granting institutions (20/217), Bachelor's-granting institutions (25/217) and Associate's-granting institutions (6/217). Most of this variation comes from within the United States - 33 of the non-PhD granting institutions are in the USA (63 total responses) whereas only 18 are outside of it (217 total responses). 

Of the courses surveyed, 137 are introductory, 79 are beyond introductory, and we have no data about one course. We discuss a split of the data by introductory and beyond introductory where appropriate in our analysis.

\begin{figure}[ht]
    \centering
    \includegraphics[width=\linewidth]{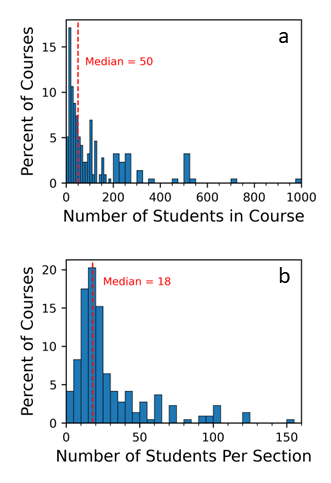}
    \caption{Number of students per course and per section. Upper histogram (a) shows the distribution of the total number of students in the course, with a median shown as a red dashed line (50) [N = 216]. Lower histogram (b) shows the distribution of the number of students per section of the course (i.e., number of students in the lab room at any time), with a median shown as a red dashed line (18) [N = 217]. }
    \label{fig:NumStuds}
\end{figure}

Next, we examine the number of students per course and the number of students per section in the course (i.e., the number of students present in the laboratory room at one time). Distributions for both of these are shown in Fig.~\ref{fig:NumStuds}. The median number of students per course is 50. Overall, there are only a few very large courses with more than 500 students. Most courses (188/216) have fewer than 200 students per course. Only three beyond introductory courses have more than 200 students. The median number of students per section is 18. Most courses (152/217) have between 10 and 40 students per section. Very few courses (14/217) have more than 75 students present in the lab at any given time; of these courses, most (12/14) are introductory, with only one beyond introductory.

\begin{figure}[ht]
    \centering
    \includegraphics[width=\linewidth]{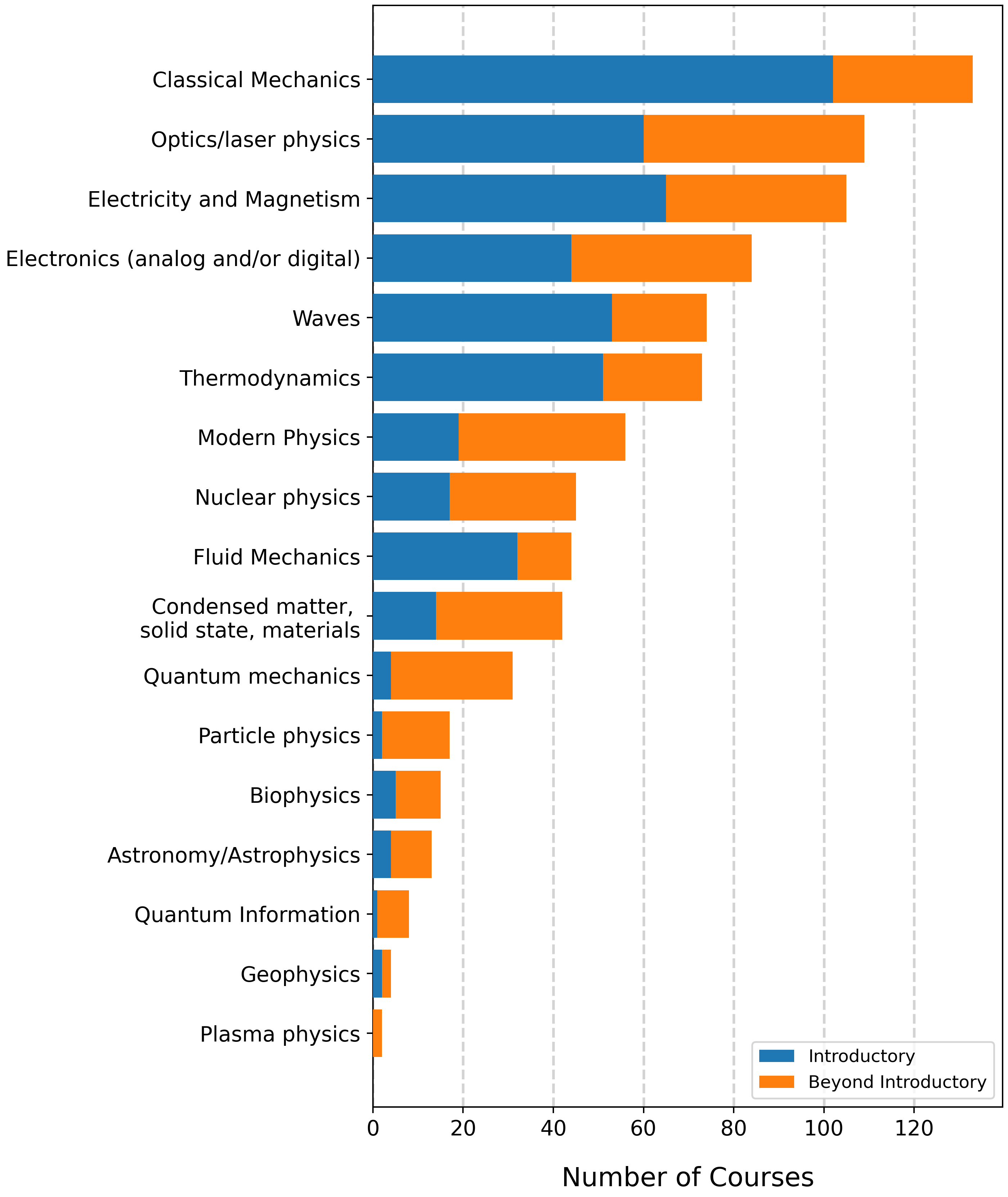}
    \caption{Topics covered in the course, split by introductory (blue, left) and beyond introductory (orange, right) courses [N = 216]. The bottom axis shows the absolute number of courses that included the topic. The most common topic for introductory courses is classical mechanics, and the most common topic for beyond introductory courses is optics and laser physics. }
    \label{fig:topics}
\end{figure}

We probe the topics covered in the course by providing a multiple-response list of possible topics and the option to type in additional topics not listed (No clear patterns emerge from the not listed responses). We present the topics covered in the courses in Fig.~\ref{fig:topics}, with a split shown between introductory and beyond introductory courses. Classical mechanics is the topic selected most often for introductory courses, while optics and laser physics was the most common topic for beyond introductory courses. More specialized areas of physics, such as plasma physics and geophysics, are rarely covered.

\begin{figure}[ht]
    \centering
    \includegraphics[width=\linewidth]{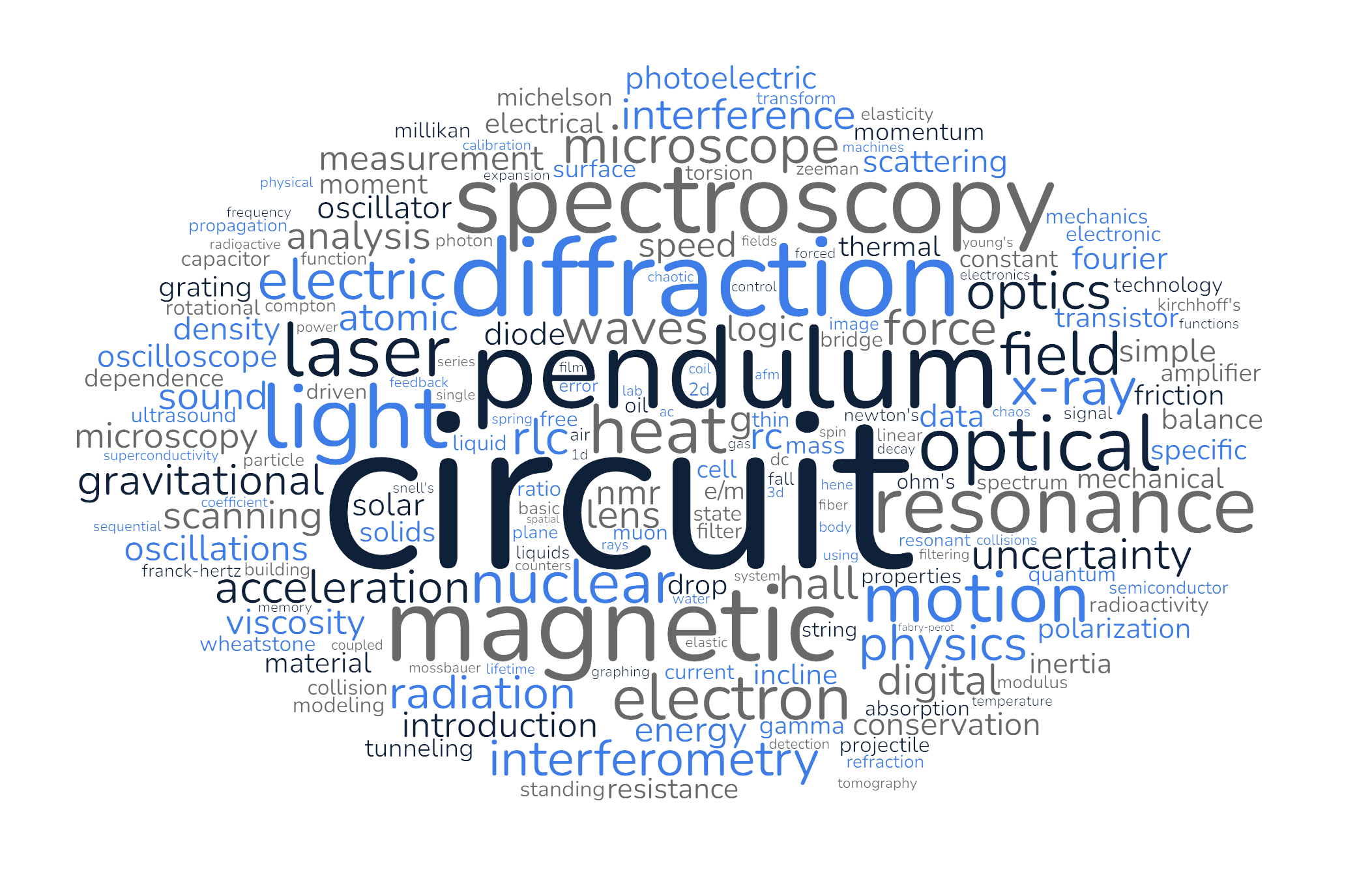}
    \caption{A word cloud showing the 200 most common words after removing stop words from the lab titles and using basic lemmatization [N = 111 courses with 1,078 lab titles]. This helps us form a visual representation of the types of experiments happening in undergraduate physics lab courses around the world. Electronics, mechanics, and optics experiments dominate the word cloud.}
    \label{fig:labtitles}
\end{figure}

In addition to the course topics, we also asked instructors to provide the titles of their lab experiments in a long-form text box. Of the 217 respondents to the survey, 111 chose to do so, and we received 1,078 lab titles from these courses. After translating all of the titles to English, we qualitatively coded them to determine the most common types of experiments occurring in undergraduate physics lab courses. These experiments are presented in Table~\ref{tab:labCodes}. This includes codes with at least 15 courses using them. Of the 1,078 lab titles received, 96 (8.9\%) were uncoded due to being too vague (e.g., classic mechanics), activities beyond a lab experiment (e.g., poster preparation), or too specific (e.g., interaction and collaboration of kirigami). The other 982 lab titles were categorized with one to three codes. The code definitions for all codes and more information about the process of coding the lab titles are located in the Appendix. Additionally, we created a word cloud of these lab titles after removing stop words~\cite{Kant} to give a visual representation of the data (See Fig.~\ref{fig:labtitles}). We hope that, as we collect more data, qualitative coding of lab titles will help in creating a taxonomy: we can work on grouping courses that complete similar types of experiments.

\begin{table}[ht]
    \centering
    \caption{Most common experiments as given by titles of lab experiments. These are the codes for all categories with at least 15 courses reporting at least one lab in this category. A total of 111 courses providing 1,078 lab titles were qualitatively coded to determine the most common experiment types. The definitions of these codes and others not shown are provided in the Appendix.}
    \label{tab:labCodes}
    \begin{tabularx}{\linewidth}{L c c } \hline \hline
         Topic & Num. & Num. Lab  \\
         & \hspace{3mm}Courses\hspace{3mm} & Titles \\ \hline
        Optics (intermediate) & 68 & 108 \\
        Kinematics & 36 & 55 \\
        Dynamics (mechanics) & 33 & 56 \\
        Electronics (intermediate) & 29 & 69 \\
        Electronics (simple) & 29 & 56 \\
        Spectroscopy & 28 & 37 \\
        Test and measurement equipment & 27 & 30 \\ 
        Thermodynamics & 26 & 56 \\
        Introduction to measurement and uncertainty & 25 & 36 \\
        Pendulum & 23 & 33 \\ 
        Optics (simple) & 23 & 39 \\
        Particle physics & 18 & 44 \\ 
        Optics (advanced) & 17 & 49 \\ 
        Advanced materials and solid state & 17 & 30 \\
        Waves & 17 & 22 \\
        Electric fields and electrostatics & 16 & 24 \\
        Fluids & 15 & 21 \\
        Magnetic fields & 15 & 18 \\
        \hline \hline
    \end{tabularx}
\end{table}

We also asked whether the laboratory course is integrated with lecture (i.e., if both are one course combined) or if the laboratory course is a separate course. There are 129 courses that are separate, while 88 are combined with a theory course. Additionally, 136 courses include lectures about statistics and/or data analysis, whereas 81 do not. 

\begin{figure}[ht]
    \centering
    \includegraphics[width=\linewidth]{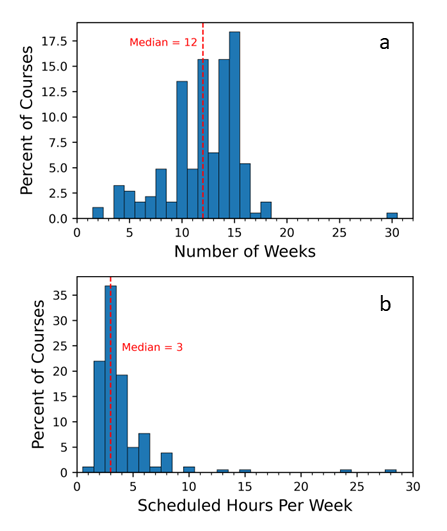}
    \caption{Number of weeks the course runs for (a) [N = 185] and number of hours per week students are scheduled to be in the lab (b) [N = 182]. Red dashed lines shown the median. The median number of weeks the course runs is 12, and the median number of hours per week is 3.}
    \label{fig:numweekshours}
\end{figure}

Next, respondents reported the number of weeks the course runs for, as well as the number of hours per week students are scheduled to be in the lab. The distributions for these are shown in Fig.~\ref{fig:numweekshours}. The median number of weeks is 12, and the median number of scheduled hours per week is three. Further, data about the number of hours per week beyond the scheduled time that students spend in the lab is presented in the Appendix; in most cases, students do not spend any time beyond what is scheduled in the lab.

\begin{figure}[ht]
    \centering
    \includegraphics[width=\linewidth]{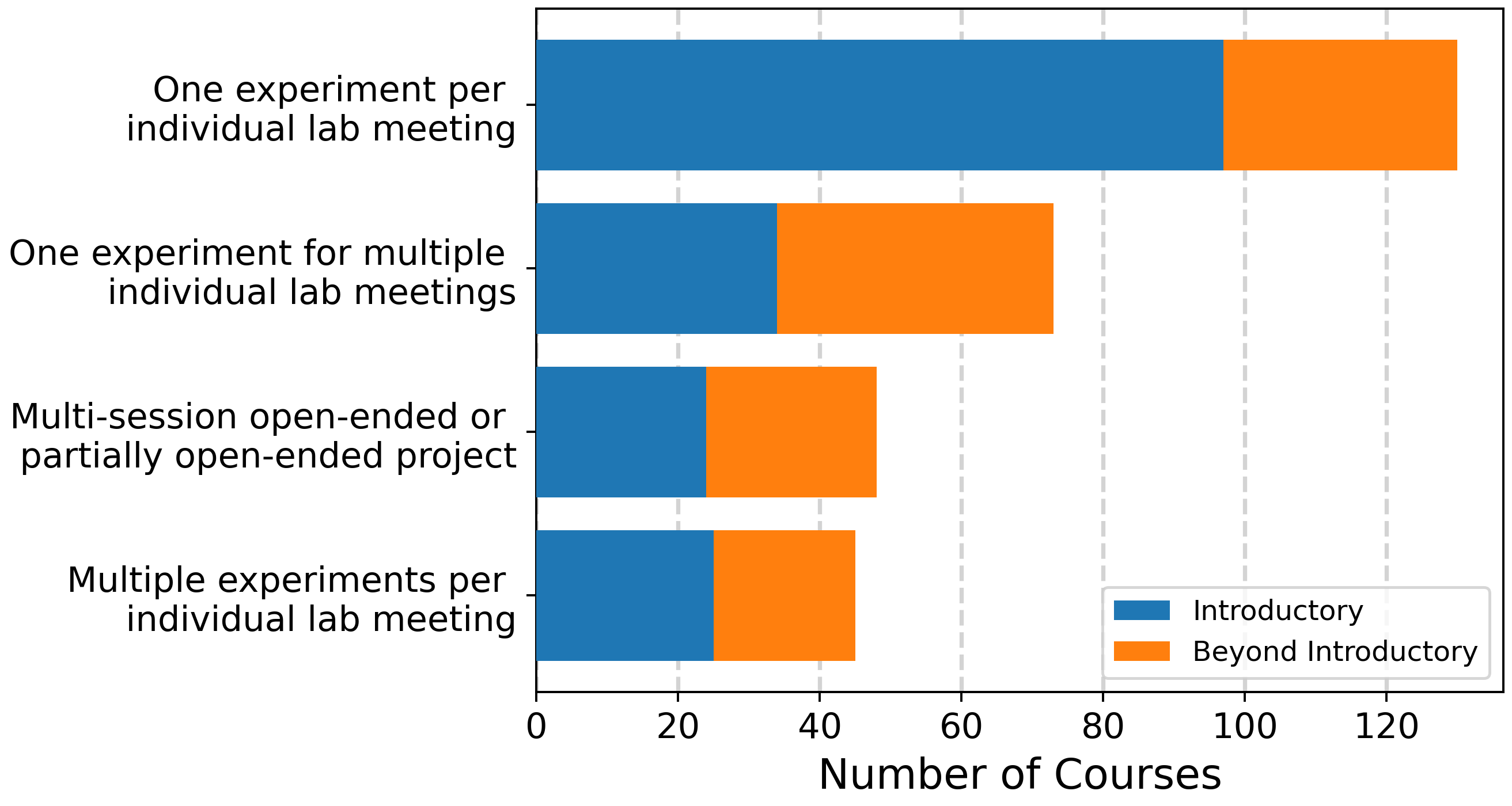}
    \caption{Types of experiments students complete in the course, split by introductory (blue, left) and beyond introductory (orange, right) [N = 216]. This question was multiple response, so respondents could select as many of these options as apply to their course. Most courses involve some component in which students complete one experiment per meeting of the course. }
    \label{fig:expttype}
\end{figure}

We further investigate the number of experiments per lab meeting students complete and if students have choice over which experiments they complete. This question is multiple response, and respondents can choose whether students complete multiple experiments per meeting, one experiment per meeting, one experiment per multiple meetings, or a multi-session open-ended project.  The distribution of responses is presented in Fig.~\ref{fig:expttype}. In most courses, students spend time doing one experiment per meeting of the course. Students are often not given a choice of which experiments they do, with 125 courses not allowing students any choice in which experiments they complete, 59 allowing students to choose their experiments for some portion of the course, and 33 allowing students to choose experiments all of the time. 

For those courses where there is a project component, we asked four additional questions about the project. There are 48 courses that contain some project component. Of these courses, students spend a median of 4.5 weeks engaged in project work. In 26 courses, students choose their project topic ``all of the time'', compared with 16 courses that allow students to choose ``some of the time'', and six courses that do not allow students to choose their own project topic. Thirty-two courses allow students to design their own project ``all of the time'', whereas 13 courses allow students to do this only ``some of the time'', and only three courses do not allow students to design their own project. Finally, in 26 courses, students always build their own experimental apparatus for the project, while in 19 courses they do this sometimes and in three courses they never do this.

\subsubsection{Students}

Respondents reported which major(s) their students have in the course, as well as the fraction of physics and astrophysics majors in the course. For the first of these questions, we presented a list of possible majors as a multiple-response question along with the option to write in majors not contained on the list. The second of these questions is multiple choice. We present the results of these items along with a split by introductory and beyond introductory in Table~\ref{tab:majors}. Because the question about the major(s) of the students is multiple response, we received many different potential groupings. The table, therefore, represents only the most common groupings (i.e., a minimum of four courses chose this grouping). Nearly a third of all courses include only physics majors, with 27\% of introductory courses and 41\% of beyond introductory courses having only physics majors. The second-most common combination overall is physics and astrophysics/astronomy majors, which accounts for another 9\% of responses. Overall, 166 out of the 217 courses surveyed (76\%) included physics majors, and therefore, 51 courses (24\%) do not include any physics majors.

Another point of discussion about majors is that the United States typically treats physics courses  differently than courses outside of the United States. Within the USA, it is very common to combine many different majors into one course at the introductory level, whereas outside of the USA, it is more common to have an introductory physics course only for physics majors, one for those training to teach high-school physics, a separate course only for engineering majors, etc. When examining only introductory courses, 81 \% (29 out of 36) introductory courses in the USA contain 0-25\% physics majors while outside of the USA, this number drops to 39\% (39 out of 101 courses).

\begin{table*}[h]
    \centering
    \caption{Most common grouping of student majors in a class and percentage of the course that is physics majors.  Only the most common groupings are shown (i.e., a minimum of four courses in that grouping).}
    \label{tab:majors}
    \begin{tabular}{lccc}\hline\hline
         & \hspace{4mm}\% Responses \hspace{4mm} & \hspace{4mm}\% Responses, \hspace{4mm} & \hspace{4mm} \% Responses, \hspace{4mm} \\ 
         &  (N = 217) & \hspace{4mm} Intro (N = 137) \hspace{4mm} & \hspace{4mm} Beyond Intro (N = 79) \hspace{4mm} \\ \hline 
        Physics &  32.2 &  27.0 &  41.8\\
        Physics, Astrophysics/Astronomy &  9.2   &  5.1   & 15.2 \\ 
        Another science (e.g., biology, geology) & 5.5 & 8.8  & 0.0   \\ 
        Physics, Physics/Astronomy teaching/pedagogy & 5.5  & 2.2 & 11.4 \\ 
        Engineering & 4.6  & 5.8 & 2.5\\
        Physics, Engineering & 2.8 & 14.6 & 5.1 \\
        Chemistry, Another science (e.g., biology, geology) & 2.3 & 3.6 & 0.0 \\
        Physics/Astronomy teaching/pedagogy & 1.8 & 2.2 & 1.3 \\
        Other & 35.9 & 43.8 & 22.3 \\ \hline
        & \% Responses & \% Responses, & \% Responses, \\ 
         &  (N = 216) & Intro (N = 136) & Beyond Intro (N = 79) \\ \hline 
        0-25\% physics majors &  34.3 & 50.0  & 7.6 \\
        25-50\% physics majors &  7.9 &  6.6   & 10.1 \\ 
        50-75\% physics majors & 2.8 & 8.1  &  6.3  \\ 
        75-100\% physics majors & 50.5 & 35.3 & 75.9 \\ 
        
        \hline \hline
    \end{tabular}
\end{table*}

Next, we provide information about how many years the students have been at the University when they take the course. Again, this question is multiple response, so instructors can choose as many options as apply to their course. These data are presented in Table~\ref{tab:studYear}. The courses in our data set lean heavily towards first-year and second-year students.

\begin{table*}[h]
   \caption{Year of students in the course [N = 216]. This question is multiple response, so instructors can select all options that apply to their course.}
   \label{tab:studYear}
   \begin{tabular}{lcc}\hline\hline
   \hspace{4mm} & \hspace{4mm}Num. Responses\hspace{4mm} & \hspace{4mm}\% Responses\hspace{4mm}\\ \hline
       1st year & 106 & 49.1\\
       2nd year & 83  & 38.4 \\ 
       3rd year & 66  & 30.6 \\ 
       4th year & 39  & 18.1 \\ 
       5th year or higher & 9  & 4.2\\ 
 \hline\hline
   \end{tabular}
\end{table*}

\subsubsection{Students working in groups}

We next inquire about the ways in which students work together in the course. Our survey results showed that 204 courses indicated students work with at least one partner, and 13 courses indicated that students work alone. Data about these 204 courses is shown in Table~\ref{tab:grouping}, including the number of lab partners, whether students stay in the same group for the entire course, and whether students choose their own groups. In most cases, students are working in pairs of their own choice and stay with this lab partner for the entire term.

\begin{table}[h]
    \centering
    \caption{Grouping of students [N = 204]. Most students work with one lab partner of their choice for the entire term.}
    \label{tab:grouping}
    \begin{tabular}{lcc}\hline\hline
        & Num.  & Percent \\
        & \hspace{2mm} Responses \hspace{2mm} & \hspace{2mm} Responses \hspace{2mm} \\ \hline
        Groups of 2 & 118 & 57.8 \\
        Groups of 3 & 63 &  30.9\\
        Groups of 4 & 19 & 9.3 \\
        Groups of 5+ & 9 &  4.4\\ 
        \hline
        Stay in same groups & 165 & 80.9\\
        Switch groups & 39 & 19.1\\ 
        \hline
        Choose their groups & 139 & 68.1\\
        Are assigned groups & 31 & 15.2\\
        Both options & 34 & 16.7 \\
        \hline\hline
    \end{tabular}
\end{table}

\subsubsection{Instructional Staff}

\begin{figure}[ht]
    \centering
    \includegraphics[width=\linewidth]{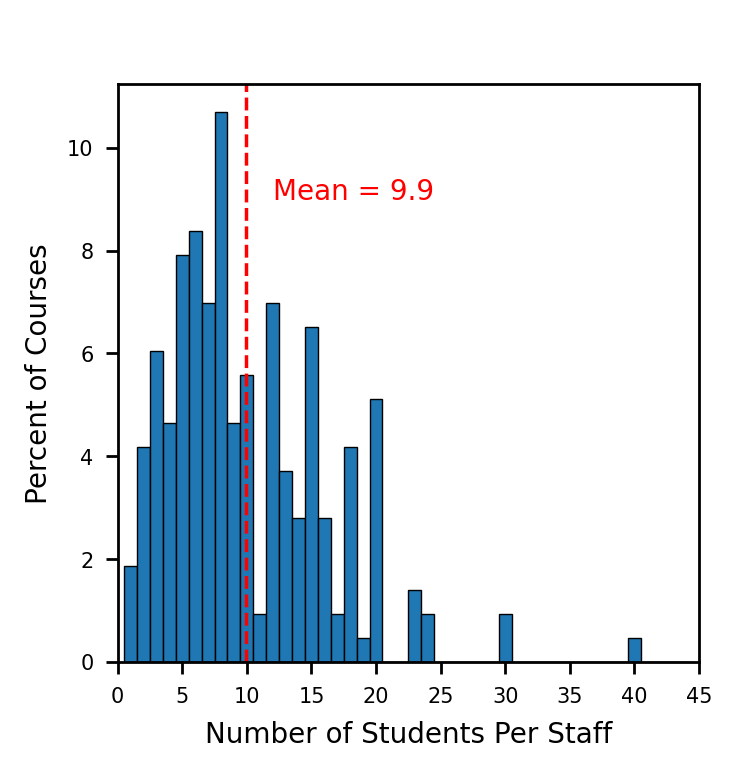}
    \caption{Distribution of the number of students per staff member in the lab room at any given time, with mean = 9.9 shown as a red dashed line.}
    \label{fig:numstudsstaff}
\end{figure}

This section of the survey asks about the number of different types of instructional staff present in the lab room with the students. Because we also know the number of students present in the lab at one time, we can determine the average number of students per staff. The means of this are shown in Table~\ref{tab:staffing}, including information about faculty members, lab technician, graduate and postdoctoral TAs, and undergraduate LAs. The distribution of the number of students per staff member is shown in Fig.~\ref{fig:numstudsstaff}. The mean number of students per staff member (after summing all possible types of staff members) is 9.9. Few courses utilize LAs, whereas many courses have faculty and TAs present with students.

\begin{table}[h]
    \centering
    \caption{Mean number of students per staff member in the lab. We pair the question probing number of staff in the room with the question about the number of students in each section to determine these averages. On average, there are a total of 9.9 students per staff member. Means take into account only courses with at least one of that type of instructional staff (e.g., courses with no undergraduate LAs are not counted in the mean number of students per LA). }
    \label{tab:staffing}
    \begin{tabular}{lcc}\hline\hline
         & Mean  & Num. \\ 
         & & \hspace{2mm}Courses\hspace{2mm} \\ \hline
        Students Per Faculty & 25 & 186 \\
        Students Per Lab Technician & 33  & 95\\
        Students Per Graduate TA & 20 & 116 \\
        Students Per Undergraduate LA  & 21  & 51\\ 
        Students Per Staff (total)  & 9.9 & 215 \\
        \hline\hline
    \end{tabular}
\end{table}

Further, respondents provided information about the frequency and type of training for both graduate TAs and undergraduate LAs. This question is multiple choice in which respondents can indicate whether training happens once per term, once per academic year, or weekly. The types of training is a multiple response question, which allows respondents to select pedagogy, grading, and familiarization with lab equipment in any combination that applies to their course. Both of these questions also have `not listed' options with the opportunity to write in a response; no patterns emerged from an analysis of these responses. These data are shown in Table~\ref{tab:trainingTypeFreq}. Nearly all courses that train TAs and/or LAs provide instruction to familiarize them with the lab equipment, while more than half also offer pedagogy or grading training. There is no standardized frequency of this training, with about one-third providing training once per term and one-quarter providing training once per academic year or weekly.

\begin{table}[h]
    \centering
    \caption{TA and LA training frequency [N = 137] and type [N = 136]. For the frequency question, respondents can type in an answer if none of the provided options capture their training schedule. Type of training is a multiple response question and also includes the ability to type in a response if a type of training is missing from the provided options.}
    \label{tab:trainingTypeFreq}
    \begin{tabular}{lcc}\hline\hline
        &  Num.  & Percent  \\
        & \hspace{2mm}Responses \hspace{2mm} & \hspace{2mm}Responses \hspace{2mm} \\ \hline
        Once per term/semester & 45 & 32.8\\
        Once per academic year &  35 & 25.5\\
        Weekly & 33 & 24.1\\
        Other & 22 & 16.1 \\ \hline

        Familiarization with equipment & 130 & 95.6\\
        Pedagogy &  80 &58.8 \\
        Grading & 80 & 58.8 \\
        Other & 7 & 5.1\\
         
        \hline\hline
        
    \end{tabular}
\end{table}

\subsubsection{Goals, Activities, and Evaluation}

\begin{figure*}[ht]
    \centering
    \includegraphics[width=\linewidth]{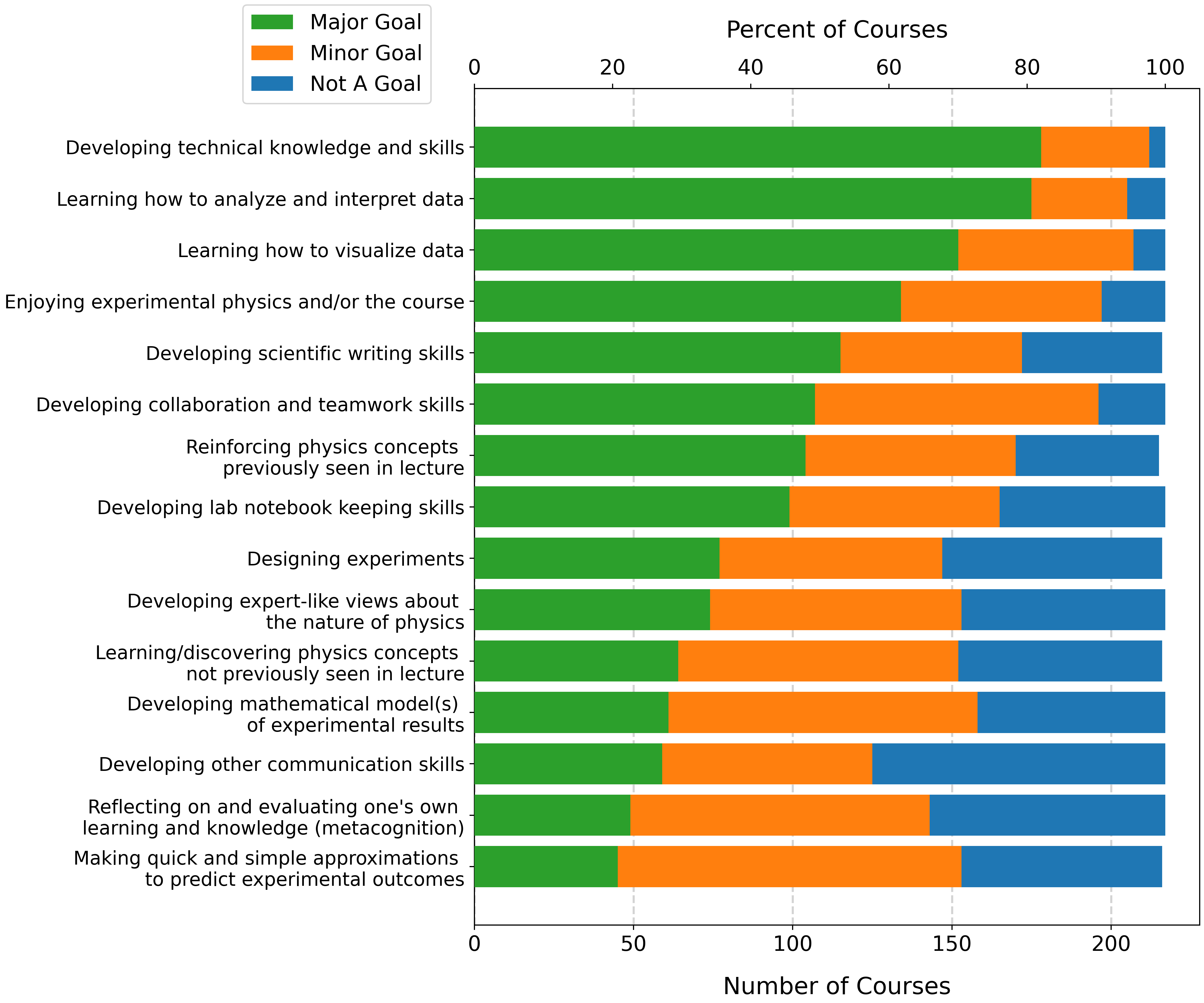}
    \caption{Course goals, split by major goal (green, left), minor goal (orange, middle), and not a goal (blue, right) for the course. Between 215 and 217 courses provided data for each goal, and the percentages are calculated using the full 217 courses for display purposes. The most commonly selected goal is developing technical knowledge and skills.}
    \label{fig:goals}
\end{figure*}

Potential course goals or learning objectives were presented as a list with options to select `Major Goal', `Minor Goal', `Not a Goal', and `Future Goal (not currently a goal)'. As previously discussed, the latter two of these categories are collapsed for all analysis. Each of the 15 goals presented had between 215 and 217 responses. A plot of the answers to this question is shown in Fig.~\ref{fig:goals}. Other course goals are possible, but there is no ``not listed'' option available for this question. The current list of goals was refined through interviews, including the addition of extra goals as requested by interviewees.

\begin{figure}[ht]
    \centering
    \includegraphics[width=\linewidth]{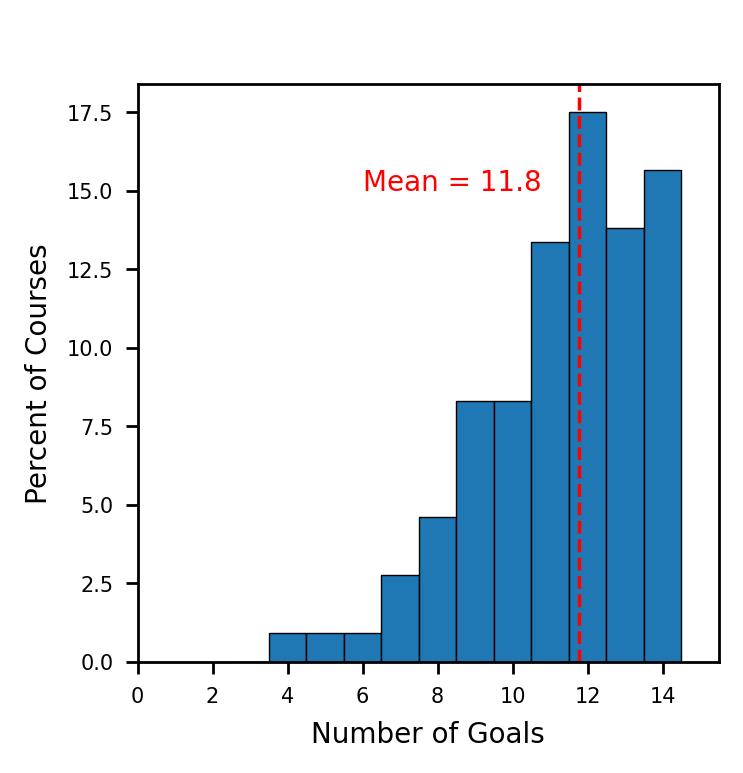}
    \caption{Distribution of the total number of goals (major plus minor) for each course. The maximum possible number of goals is 15 (all of the provided course goals). The mean number of goals per course (red dashed line) is 11.8.}
    \label{fig:numgoals}
\end{figure}

We also examine the total number of goals (major plus minor) selected for each course.  This distribution is shown in Fig.~\ref{fig:numgoals} and the mean is 11.8 goals out of the possible 15. On average, courses have 6.9 major goals and 4.9  minor goals. There are no significant differences in number of course goals (either total, major, or minor) for introductory and beyond introductory courses.

The first question in the activities section asks about whether a specific branded instruction technique is used (such as modeling instruction, SCALE-UP, or ISLE). Most courses (135/212) do not use any type of branded instruction method. Similarly, a question about RBAIs reveals that most courses (148/212) do not use any of these.

\begin{figure}[ht]
    \centering
    \includegraphics[width=\linewidth]{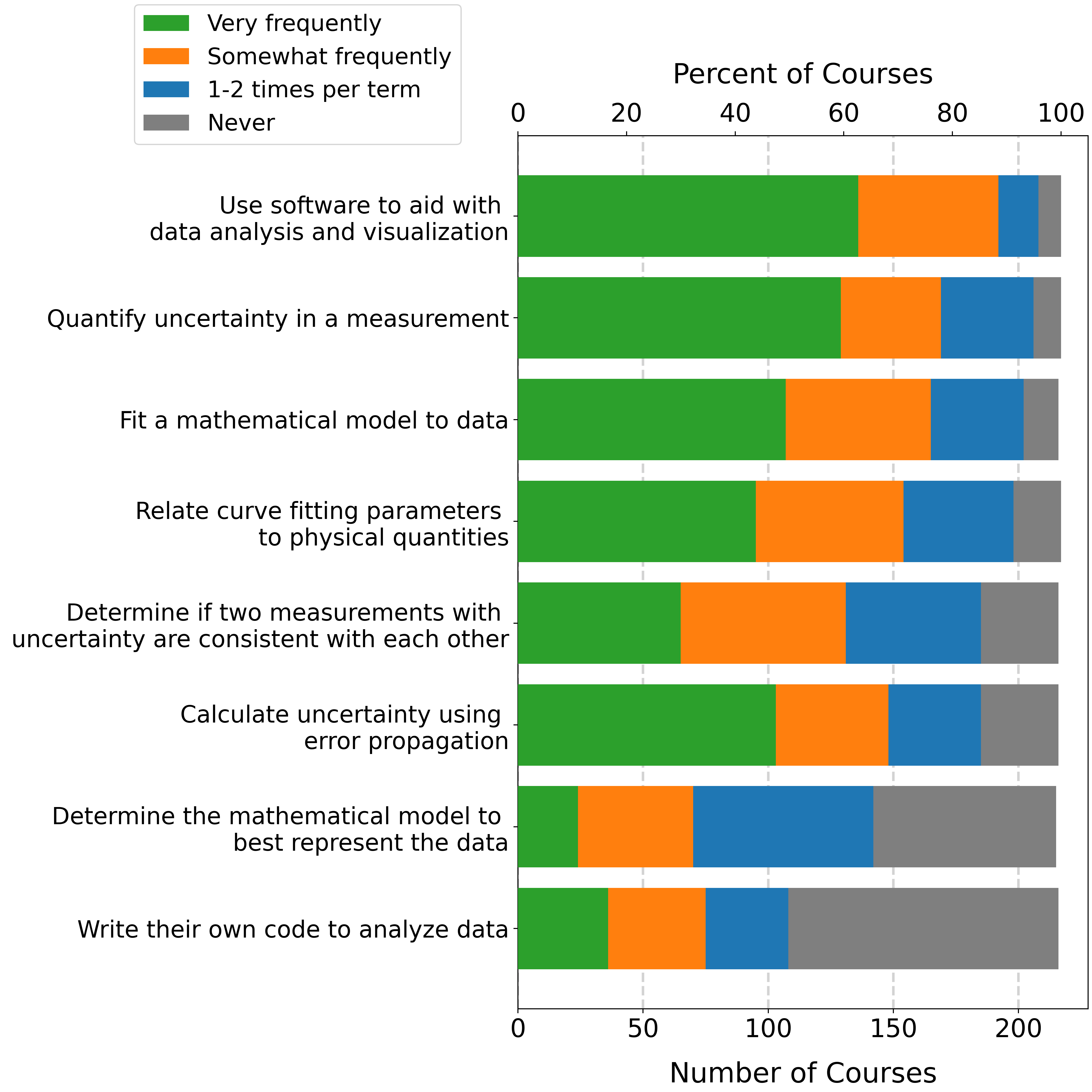}
    \caption{Courses with data analysis activities [N = 215 - 217]. Bars represent number of courses (bottom axis) and percent of courses (top axis) that include various activities related to analyzing data, such as error propagation and curve fitting. Bars are split based on frequency of the activity - very frequently (left, green), somewhat frequently (second from left, orange), 1-2 times per term (second from right, blue), and never (right, gray). In general, students participate in each of the data analysis activities to some extent in at least half of all courses. The least popular activity was for students to write their own code to analyze data.}
    \label{fig:act-dat}
\end{figure}

\begin{figure}[ht]
    \centering
    \includegraphics[width=\linewidth]{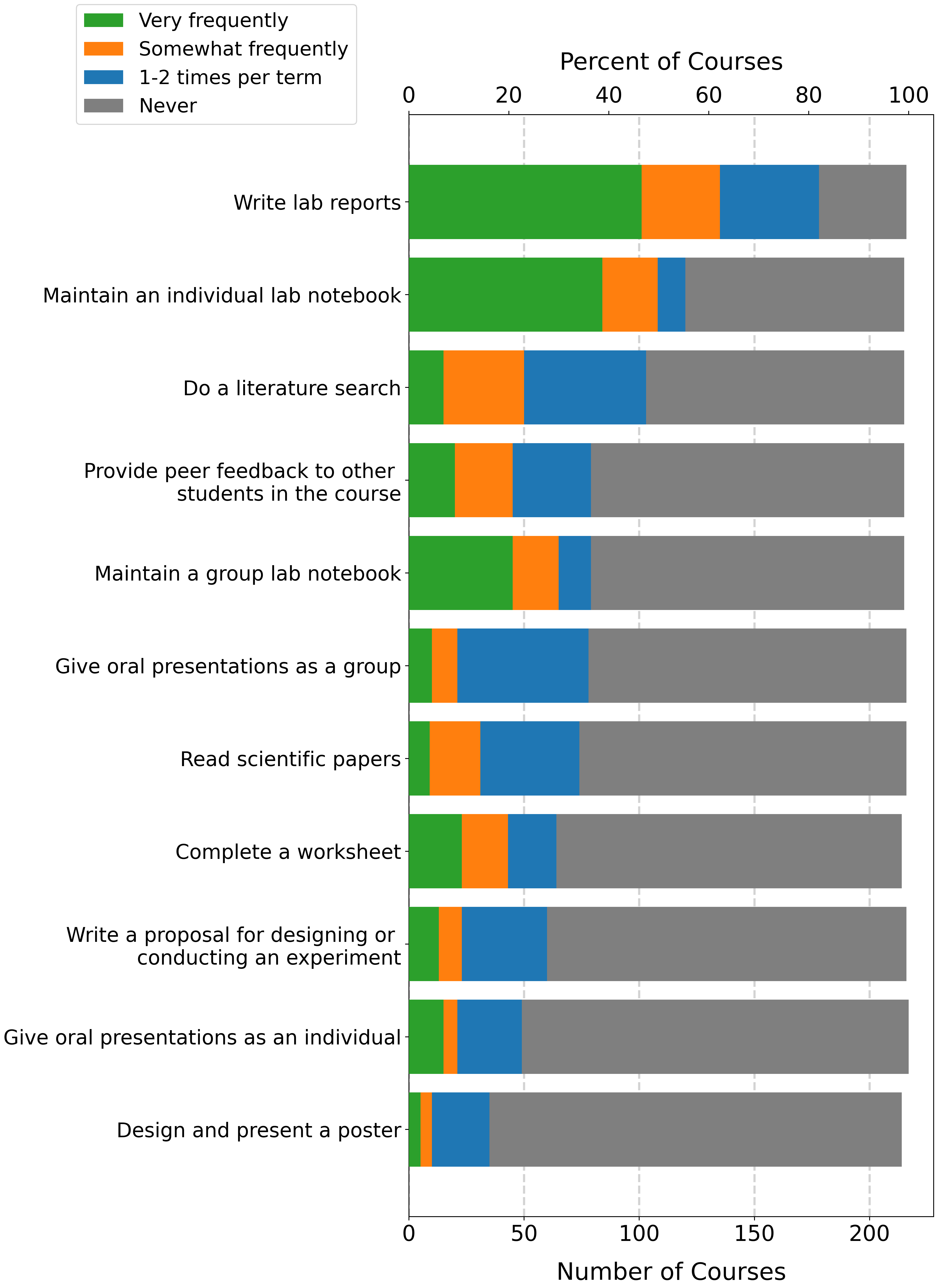}
    \caption{Courses with communication activities [N = 214 - 217]. Bars represent number of courses (bottom axis) and percent of courses (top axis) that include various activities related to communication skills, such as peer feedback and lab reports. Bars are split based on frequency of the activity - very frequently (left, green), somewhat frequently (second from left, orange), 1-2 times per term (second from right, blue), and never (right, gray). Writing lab reports is a common activity, with more than 80\% of courses indicating that student engage in this to some extent. Very few courses have students present posters, give presentations, or write proposals for experiments.}
    \label{fig:act-comm}
\end{figure}

\begin{figure}[ht]
    \centering
    \includegraphics[width=\linewidth]{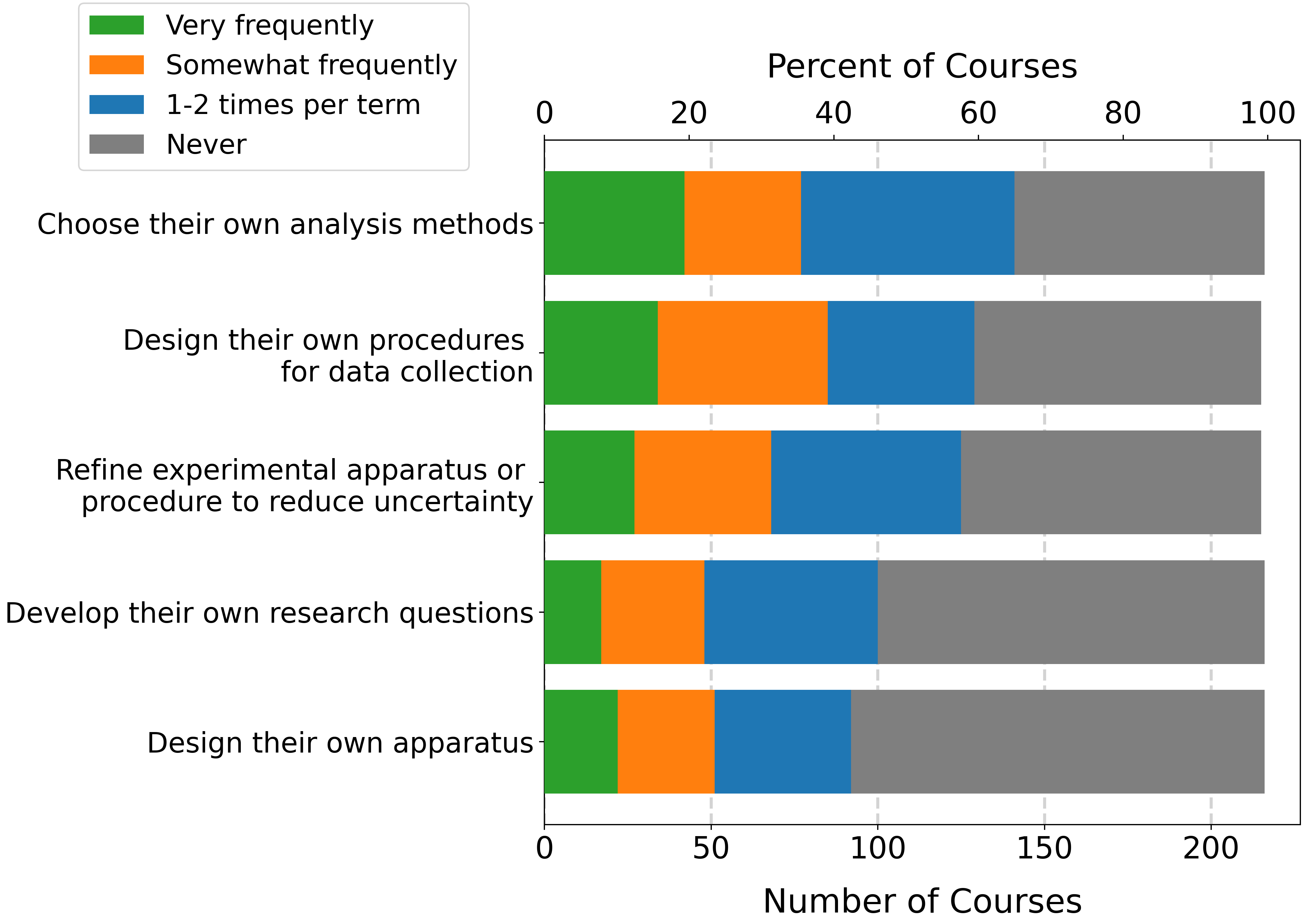}
    \caption{Courses with student decision-making activities [N = 215 - 217]. Bars represent number of courses (bottom axis) and percent of courses (top axis) that include various activities related to decisions made by students, including choosing their own procedures and analysis methods. Bars are split based on frequency of the activity - very frequently (left, green), somewhat frequently (second from left, orange), 1-2 times per term (second from right, blue), and never (right, gray). Students engage in these activities in many courses, although in most cases, they are only doing these 1-2 times per term as opposed to other activity categories that have more responses in the very frequently category.}
    \label{fig:act-deci}
\end{figure}

\begin{figure}[ht]
    \centering
    \includegraphics[width=\linewidth]{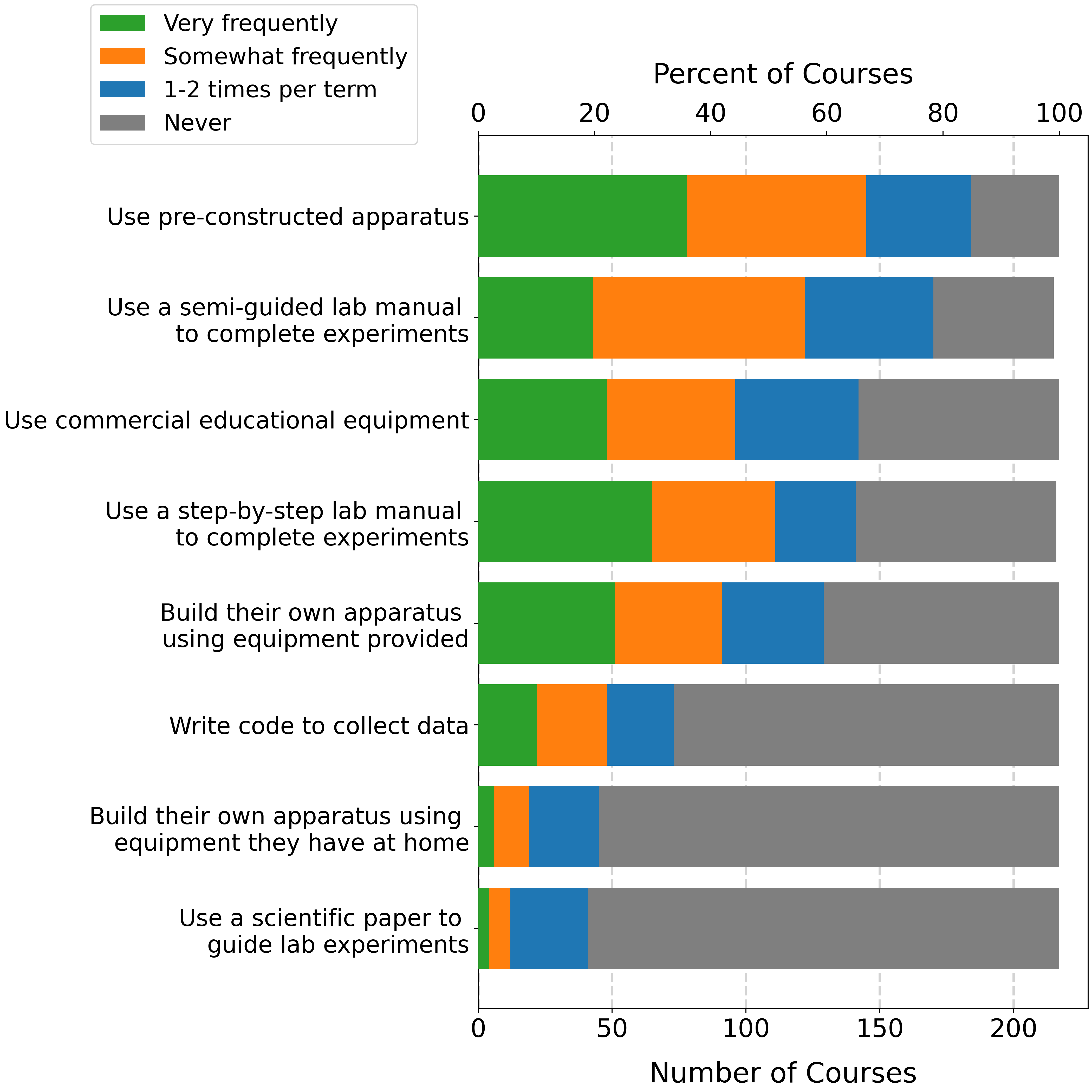}
    \caption{Students' engagement with materials in lab courses [N = 215 - 217]. Bars represent number of courses (bottom axis) and percent of courses (top axis) that include various methods of student interaction with materials, including use of commercial equipment (such as PASCO or TeachSpin), as well as students building their own apparatus. Bars are split based on frequency of engagement with the activity - very frequently (left, green), somewhat frequently (second from left, orange), 1-2 times per term (second from right, blue), and never (right, gray). }
    \label{fig:act-mat}
\end{figure}

\begin{figure}[ht]
    \centering
    \includegraphics[width=\linewidth]{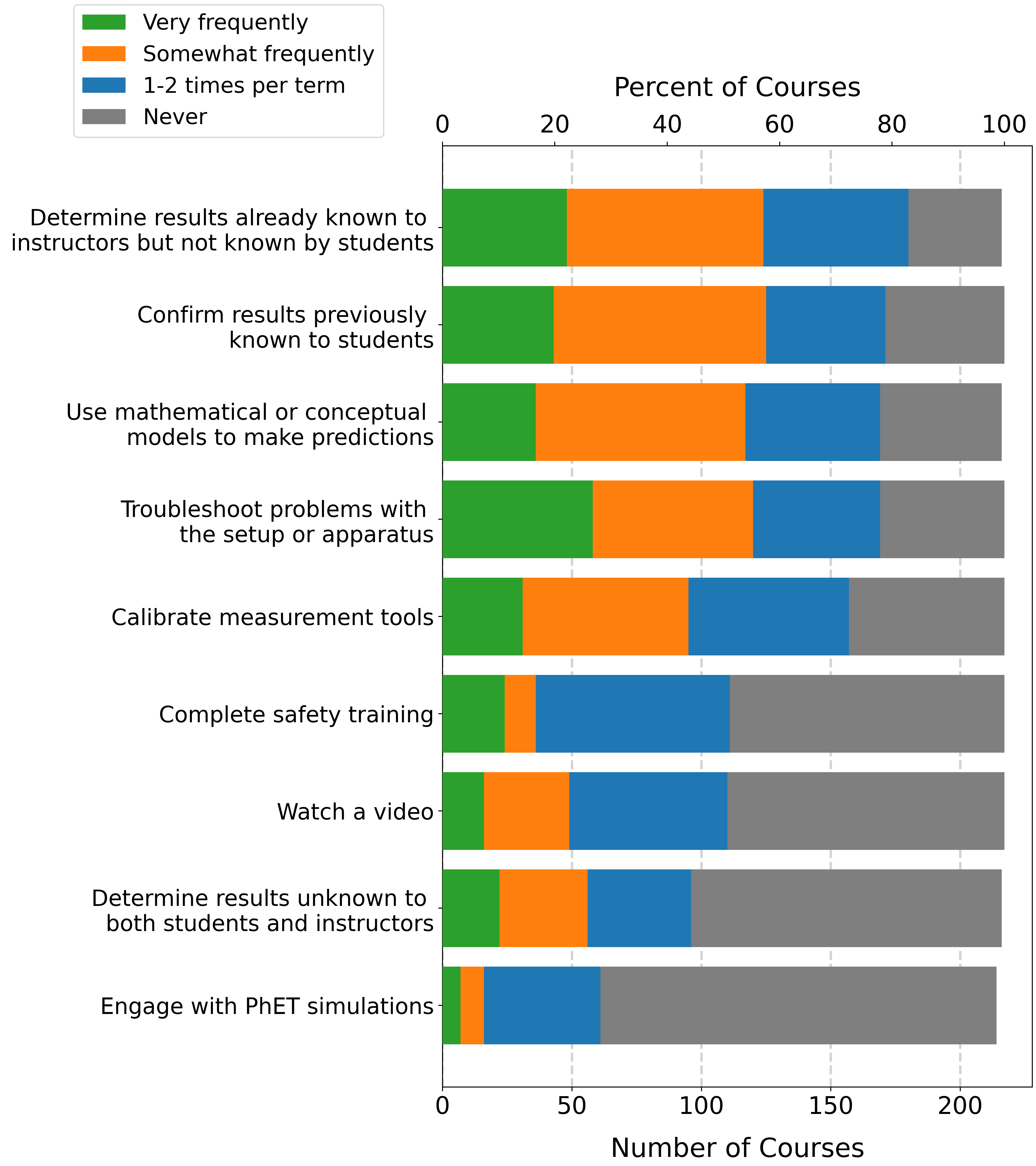}
    \caption{Courses with modeling and other activities [N = 214 - 217]. Bars represent number of courses (bottom axis) and percent of courses (top axis) that include various activities related to modeling, such as using models to make predictions, as well as other activities such as watching a video or completing a safety training. Bars are split based on frequency of engagement with the activity - very frequently (left, green), somewhat frequently (second from left, orange), 1-2 times per term (second from right, blue), and never (right, gray).}
    \label{fig:act-mod}
\end{figure}

Next, we present respondents with a list of 41 possible activities broken up into five categories --- data analysis, communication, student decision-making, materials, and modeling/other activities. Plots of the responses to the Likert-style questions for each of these categories are shown in Figs.~\ref{fig:act-dat}, ~\ref{fig:act-comm}, ~\ref{fig:act-deci}, ~\ref{fig:act-mat}, and ~\ref{fig:act-mod}. These figures are broken down by Likert response (very frequently, somewhat frequently, 1-2 times per semester/term, and never, where we have again collapsed an aspirational scale point with never). We find that courses engage in a wide variety of activities. In some cases, such as in the collection of activities relating to both data analysis and student decision-making, at least half of the courses selected that they participated in all activities to some extent. The split between the frequencies students engage in activities also generally occurs as expected. For example, students typically write lab reports very frequently, but design and present a poster 1-2 times per term. The student decision-making category, in particular, has a large number of activities with responses of 1-2 times per term.

Further, we find that students rarely use a scientific paper to guide their lab experiments - in most cases they either use a step-by step lab manual or a semi-guided lab manual. In more than 80\% of courses, students use a pre-constructed apparatus to some extent. Further, in most courses (more than 80\%), students spend some time determining results already known to the instructor, but not yet known to the students, though in nearly 80\% of courses, students engage with activities where they are confirming results they have already learned in a lecture course.

A list of potential items that might be graded for inclusion in a student's final course grade is presented to respondents, and they are able to select all of the ones they use in their own course to assign student grades (multiple response). This list of 18 potential things might not fully span the space of items included in a grade, and so ``not listed'' with an option to write in other items is included in the survey. Table~\ref{tab:grades} shows the number of courses that include each option. The analysis of ``not listed'' write-in responses did not reveal any patterns. Most courses (about 75\%) use lab reports to assign grades to students, with attendance and participation being the second most common item with more than half of the courses using this.

\begin{table}[ht]
    \centering
    \caption{Items included in final course grade [N  = 216]. This multiple response item allows respondents to select multiple items, as well as an option to write in anything not listed. }
    \label{tab:grades}
    \begin{tabular}{lcc}\hline\hline
         & Num. & Percent \\ 
         & \hspace{2mm}Responses\hspace{2mm} & \hspace{2mm}Responses \hspace{2mm}\\ \hline
        Lab report & 160 & 74.1\\
        Attendance/participation & 117 & 54.2\\
        Lab notebooks & 99 & 45.8\\
        Accuracy/precision of results & 88 & 40.7\\
        Oral presentation & 69 & 31.9\\ 
        Observation of students & 62 & 28.7 \\
        Prelab calculations & 59 &  27.3\\
        Written exam & 51 & 23.6\\
        Interview/meeting after the lab & 40 & 18.5 \\
        Worksheets & 33 & 15.3\\
        Partial lab report & 31 & 14.4\\
        Prelab measurement/analysis plan & 29 & 13.4 \\
        Practical exam & 27 & 12.5\\
        Quiz/interview prior to working & 24 & 11.1\\ 
        Poster presentation & 22 & 10.2\\
        Prelab quiz & 22 & 10.2\\
        Peer feedback & 19 & 8.8\\
        Prelab video &  18 & 8.3\\
        Not Listed & 31 & 14.4\\
        \hline\hline
    \end{tabular}
\end{table}

Finally, we can combine goals, activities, and items graded to determine how instructors are attempting to meet their course goals as discussed previously.  We present the results of this analysis in Table~\ref{tab:matching}. In this analysis, we collapse the categories of major goal and minor goal together.

We find that courses typically engage in a higher percentage of activities related to a goal than items graded related to that goal; this percentage is often much larger (in some cases, more than four times). There is no correlation between having a certain goal for the course and the percentage of activities related to that goal or the percentage of items graded related to that goal, thus showing that instructors take many different paths in attempting to achieve their course goals.

\begin{table*}[ht]
    \centering
    \caption{Matching of goals with activities and items graded. The number of courses represents those who selected the goal as either a major or minor goal for their course. This table shows, if an instructor selects a goal, how many activities and items graded they have selected on average (mean) that match that goal. These are shown as fractions as well as percentages. The fractions allow visualization of the total number of matched activities and items graded for each goal, while the percentages allow for comparison between these matched items more easily. One goal (enjoyment of experimental physics and/or the course) is not shown in this table because no activities or items graded match that goal. Additionally, the goal related to approximations has no items graded matched with it (though it does have matched activities). In general, we find that instructors have a higher percentage of activities for a specific goal than items graded for that goal.}
    \label{tab:matching}
    \begin{tabular}{m{2.9in} c|cc|cc}\hline\hline
 & &\multicolumn{2}{c|}{Activities} & \multicolumn{2}{c}{Items Graded}\\
 & \hspace{2mm}Num. Courses\hspace{2mm} & \hspace{3mm}Fraction\hspace{3mm} & \hspace{3mm}Percent\hspace{3mm} & \hspace{3mm}Fraction\hspace{3mm} & \hspace{3mm}Percent\hspace{3mm} \\ \hline \vspace{2mm}
  Developing mathematical model(s) of experimental results & 158 & 5.7/7 & 81 & 1.8/5 & 36 \\ 
 \vspace{2mm}
  Making quick and simple approximations to predict experimental outcomes & 153 & 1.6/2 & 80 & N/A & N/A \\ 
 \vspace{2mm}
  Learning how to analyze and interpret data & 205 & 8.4/11 & 76 & 1.6/5 & 32 \\ 
 \vspace{2mm}
 Reinforcing physics concepts previously seen in lecture & 170 & 7.0/10 & 70 & 2.8/10 & 28 \\ 
 \vspace{2mm}
  Developing scientific writing skills & 172 & 1.3/2 & 65 & 1.2/3 & 40 \\ 
 \vspace{2mm}
 Learning physics concepts not previously seen in lecture & 152  & 9.1/14 & 65 & 2.2/6 & 37 \\ 
 \vspace{2mm}
  Learning how to visualize data & 207 & 2.5/4 & 63 & 1.4/4 & 35 \\ 
 \vspace{2mm}
  Developing expert-like views about the nature of the process of doing experimental physics & 153  & 18.2/30 & 61 & 2.4/8 & 30 \\ 
 \vspace{2mm}
  Designing experiments & 147 & 5.8/10 & 58 & 0.59/2 & 30 \\
 \vspace{2mm}
  Developing lab notebook keeping skills & 165 & 1.1/2 & 55 & 0.57/1 & 57 \\ 
 \vspace{2mm}
 Developing technical knowledge and skills & 212  & 3.2/6 & 53 & 1.4/4 & 35 \\ 
 \vspace{2mm}
  Reflecting on and evaluating one's own learning (metacognition) & 143 & 0.44/1 & 44 & 0.10/1 & 10 \\ 
 \vspace{2mm}
  Developing collaboration and teamwork skills & 196 &1.2/3 & 40 & 0.66/2 & 33 \\ 
 \vspace{2mm}
  Developing other communication skills & 125 & 2.0/5 & 40 & 1.1/5 & 22 \\ \hline \hline

    \end{tabular}
\end{table*}

\section{Conclusions and Future Research}
\label{sec:conclusion}
 
Here, we presented the development of a survey designed to collect data that will allow us to create a taxonomy of lab courses with additional data collection. The goals of this paper were to detail the development and validation of this survey, as well as to present initial findings from the data collected. We detailed the steps of developing the survey, including the initial brainstorming sessions, collection of information according to the Spinnenweb categories~\cite{VanDenAkker2003}, and organization of the questions into the final version of the survey. Further, we performed interviews with lab instructors in order to determine that we have evidence of validity of the survey. We find evidence for construct, content, and face validity through both the development of the survey, as well as via the interviews. We analyzed survey results, including responses from 217 courses in 41 countries, to present an initial view of physics lab courses worldwide.

This global study found many similarities between these courses --- for example, in almost all courses, students work with at least one partner. Additionally, a goal of nearly all courses is for students to develop technical knowledge and skills. We also found many differences between these courses, such as the number and types of goals of the course, the activities students participate in, and the student-to-staff ratio. Further, we determined that there is no significant difference between introductory and beyond introductory courses in terms of the number of course goals they have. We also showed that in terms of data analysis and student-decision making, at least half of all courses participate in some extent in all activities in these categories.

We also presented interesting results about the level of guidance provided to students, especially that students are much more likely to use pre-constructed apparatuses (rather than building their own) and are often engaging in activities that confirm results already learned in a lecture course. More research on open-ended lab courses and how they differ from traditional courses could be useful for understating the spectrum of guidance students receive in lab courses. 

The results of this study have also raised several new questions in investigating undergraduate lab courses. For example, we find that students tend to work with other students in these courses, but we do not explicitly know why courses are structured in this way. It might be due to equipment limitations, logistical constraints due to two or more people being necessary to actually perform the experiment, pedagogical reasons, or perhaps simply tradition. We frequently assume that working collaboratively has pedagogical benefits, but this study does not investigate those, though group work in lab courses is addressed in other PER literature~\cite{Werth2022b, Sundstrom2022, Stump2023, Dew2024, Pols2023b}. Further, we have no information about how well instructors are meeting their course goals. Because the average number of goals per course is so high (mean = 11.8), it seems unlikely that instructors can focus equal time to all of these goals. Even considering major goals only, courses have a mean of 6.9 major goals, which is a significant number of goals for a short course that might only meet a few hours per week for a term. We have no detailed information about actions instructors take in order to meet these goals aside from knowing some of the activities and graded items that might relate to these goals. Thus, the initial results of this survey suggest many ideas for future PER in lab courses.

We hope to continue data collection in order to make more claims based on an expanded data set and eventually build a taxonomy of laboratory courses in order to help classify these courses and make comparisons easier. We need at least an order of magnitude more data in order to accomplish this goal, as we would eventually like to use a clustering analysis in order to analyze the data and find clusters that represent different types of courses~\cite{Ding2009, Everitt2011}.

Further, if we collect a significant amount of data in individual countries, we would like to analyze the landscape of undergraduate physics laboratory courses in these specific countries; this would require many more responses from each individual country.

Finally, with more data, we can present a more complete view of the landscape of undergraduate physics lab courses worldwide to give instructors and researchers a broad perspective as they work to improve physics laboratory instruction globally.

\section{Acknowledgements}

The authors thank Rachael Merritt for her contributions in helping match goals, activities, and items graded. We also would like to thank the 23 instructors who participated in interviews for the generosity of their time and invaluable help in developing our survey, as well as the course instructors who filled out the survey once it was finished. We would like to thank the Imperial College London European Partners Fund which supported the workshop where the need for this work was first identified. This work is supported by NSF DUE 1914840, DUE 1913698, and PHY 2317149 .

\clearpage

\bibliography{bib.bib}

\clearpage

\appendix

\renewcommand\thefigure{\thesection\arabic{figure}}    
\renewcommand{\thetable}{\thesection\arabic{table}}

\section{Additional Survey Results}
\label{sec:appendix}
\setcounter{figure}{0}  
\setcounter{table}{0}

In this appendix, we present a list of countries represented in the interviews and additional results from the survey.

Interviews took place with instructors from the following countries: Australia, Brazil, Canada, Chile, China, Colombia, Finland, France, Georgia, Greece, India, Indonesia, Ireland, Israel, Kenya, Norway, Oman, Pakistan, Poland, South Africa, South Korea, and the United States. Each country had one instructor interview except for the United States, which had two (the first was to check the interview protocol as well as collect data; the USA is varied enough to warrant two interviews). Because our authors are from Germany, Italy, England, and the Netherlands, we specifically did not contact instructors in these countries for interviews because our authors could provide the necessary information.

 Table~\ref{tab:countryRespondents} shows the number of responses to the survey we received for each country.

\begin{table*}[h]
   \caption{Respondents by Country. The number of unique courses per country that are included in the final data set (N = 217), as well as the percent of responses from each country are listed. The countries where the authors are from have a higher than average percentage of responses.}
   \label{tab:countryRespondents}
   \begin{tabular}{lcc}\hline\hline
   Country\hspace{4mm} & \hspace{4mm}Num. Responses\hspace{4mm} & \hspace{4mm}\% Responses\hspace{4mm}\\ \hline
       United States & 63 & 29\\
       Germany & 33  & 15 \\ 
       Italy & 17  &7.8\\ 
       United Kingdom & 11  &5.1\\ 
       Netherlands & 10  &4.6\\ 
       Canada & 6  &2.8\\
       Hong Kong & 6  &2.8\\
       Austria & 5  & 2.3 \\
       Finland & 5  &2.3 \\
       Slovenia & 5  &2.3 \\
       Spain & 5  & 2.3 \\ 
       Belgium & 4  &1.8 \\ 
       Latvia & 4  &1.8\\
       Uruguay & 4  & 1.8\\ 
       China & 3  &1.4\\
       Switzerland & 3  & 1.4\\
       Czech Republic & 2  & 0.92\\
       France & 2  & 0.92\\
       Kenya & 2  & 0.92\\
       India & 2  & 0.92\\
       New Zealand & 2  & 0.92\\
       Pakistan & 2  & 0.92\\
       Portugal & 2  & 0.92\\
       Taiwan & 2  & 0.92\\
       Argentina & 1   & 0.46\\
       Australia & 1  & 0.46 \\
       Brazil & 1  & 0.46\\
       Bulgaria & 1  & 0.46\\
       Chile & 1  & 0.46\\
       Colombia & 1  & 0.46\\
       Ecuador & 1  & 0.46\\
       Greece & 1  & 0.46\\
       Mexico & 1  & 0.46\\
       Norway & 1  & 0.46\\
       Oman & 1  & 0.46\\
       Poland & 1  & 0.46\\
       Slovakia & 1  & 0.46\\
       South Africa & 1  & 0.46\\
       Thailand & 1  & 0.46\\
       Vietnam & 1  & 0.46\\
       Zimbabwe & 1  & 0.46\\
\hline\hline
   \end{tabular}
\end{table*}

Next, we present in Table~\ref{tab:beyondScheduled} the number of hours per week beyond the scheduled time that instructors estimate students spend in the lab.

\begin{table*}[h]
   \caption{Number of hours per week beyond the scheduled time students spend in the lab [N = 185]. In most courses, students do not spend any time in the lab other than what is scheduled.}
   \label{tab:beyondScheduled}
   \begin{tabular}{lcc}\hline\hline
   \hspace{4mm} & \hspace{4mm}Num. Courses\hspace{4mm} & \hspace{4mm}\% Responses\hspace{4mm}\\ \hline
       0 hours & 103 & 56\\
       1 - 3 hours & 56  & 30 \\ 
       4 - 6 hours & 7  & 3.8\\ 
       More than 6 hours & 6  & 3.2\\ 
       Unknown & 4  & 2.2\\ 
       Not allowed & 9  & 4.9\\
  
\hline\hline
   \end{tabular}
\end{table*}

Finally, we provide the definitions of all codes used to qualitatively code the lab titles in Table~\ref{tab:codeDefs}. Codes were developed emergently, where each lab title was first categorized on  fine-grained scale and then codes were collapsed to create the final categories. In some cases, titles might be double- or triple-coded as they fall into two or three clear categories. For example, positron emission tomography (PET) is both a particle physics experiment but also one with medical applications and therefore is coded in both categories.

\clearpage 
\begin{longtable*}[ht]{>{\raggedright\arraybackslash}p{1.3in} p{3.3in} c c}

\caption{Definitions of codes used to qualitatively code the lab titles as well as the number of courses and number of lab titles coded for each. We include in some cases specifically items which are not included as part of the code.} 
\label{tab:codeDefs}
\\ \hline \hline 

    Code & Definitions & Num.  & Num. Lab \\ 
    & & Courses & Titles \\ \hline
         
Advanced materials and solid state &  crystals (including 2D crystals), ferrite hysteresis, fluorophore characterization, magnetic hysteresis, nanoparticles, photovoltaics, plasmon resonance, PN junctions, quantized conduction, quantum dots, quantum Hall effect, semiconductors, solar power, superconducting quantum interference device (SQUID), superconductivity, surface physics, surface roughness via advanced microscopy techniques, thermionic emission, tribology  & 17 & 30 \\  \hline
 
Arduino &   & 3 & 3 \\  \hline

Blackbody radiation & blackbody radiation, Planck radiation, thermal radiation  & 5  & 5  \\  \hline

Charge-to-mass ratio of electron &   & 8 & 8\\  \hline

Density Measurement & Archimedes' principle, measuring the density of liquids and/or solids  & 8 & 9\\  \hline

Dynamics (mechanics) & Atwood machine, collisions, conservation laws (energy, momentum), energy, drag, forces, friction, Maxwell wheel, measuring gravitational constant G (NOT measuring gravitational acceleration g), Newton's laws, orbits, torque, work; NOT pendulum, NOT springs   & 33 & 56 \\  \hline

Electric fields/electrostatics & 2D electric potential [V(x,y)], capacitance, Coulomb's law, current balance, dielectric properties of materials, forces between capacitor plates  & 16 & 24 \\  \hline

Electron diffraction &    & 6 & 6 \\  \hline

Electronics (advanced) & adders, central processor creation, counters, decoders, digital circuits, digital microchips, drivers for hardware devices, electronic oscillator, flip-flops, Fourier transform, harmonic oscillator circuit implementation, logic gates, multiplexors, Nyquist-Shannon sampling theorem, registers, serial adders, small radio construction, stopwatch with OLED display, triodes  & 10 & 30\\  \hline

Electronics (intermediate) & chaotic circuits, coaxial transmission line, diodes, electric engines, electronic feedback and/or control, electronic hysteresis, impedance, IV characteristics (NOT Ohm's law), light-emitting diodes (LEDs), lock-in amplifier, magnetoresistive effects, motors, nonlinear circuits, operational amplifiers (op-amps), passive filters, power factor measurement, RF spectrum analyzer, RLC circuits, Thevenin circuits, toggle circuits, torsion magnetometer, transients, transistors, Wheatstone bridge  & 29 & 69 \\  \hline

Electronics (simple) & AC circuits, capacitors, DC circuits, inductors, internal resistance, Kirchoff's laws,  material resistivity and/or wire resitivity, Ohm's law, resistors, RC circuits, RL circuits,  series and parallel circuits, voltage sources   & 29 & 56 \\  \hline

Fluids & aerodynamics, Bernoulli's equation, Brownian motion, diffusion, fluid flow, Hagen-Poiseuille's law, liquids, rheological behavior, stable Kaye effect, Stokes law, superfluid helium, surface tension   & 15 & 21 \\  \hline

Franck-Hertz Experiment &   & 6 & 6 \\  \hline

Hall Effect &   & 10 &  11 \\  \hline

Interferometry & Fabry-Perot, Mach-Zehnder, Michelson  & 12 & 14 \\  \hline

Introduction to measurement and uncertainty & generic "introduction to equipment" or "introduction to measurement", control of variables, statistics lectures, uncertainty analysis and/or error propagation  & 25 & 36 \\  \hline

Kinematics  & center of mass, free fall, gyroscope, inclined plane, measuring gravitational acceleration, g (NOT gravitational constant, G, and NOT the use of a pendulum), moment of inertia, motion analysis, parabolic motion, projectile motion, rotational motion  & 36 & 55 \\  \hline

Lasers & diode laser, fiber laser, laser pulses, Nd:YAG laser; NOT HeNe lasers, NOT laser spectroscopy  & 11 & 12 \\  \hline

Magnetic fields  & Earth's magnetic field,  eddy currents, Faraday's law and/or induction, Helmholtz coils, induced electromotive force, solenoids  & 15 & 18 \\  \hline

Magnetism  & force-distance relationship, Lorentz force, magnetic domains, magnetic force on conductor   & 11 &  12 \\  \hline

Materials (simple)  & anelasticity of solids, bending a bar, deformation, elasticity, elastic torsion,    elongation of a wire, plasticity of solids, Young's modulus  & 10 & 14  \\  \hline

Mechanical oscillations  & coupled oscillators, damping, forced mechanical oscillator (with and without friction), harmonic motion, mass/spring, normal modes, resonance    & 12 & 16 \\  \hline

Medical applications  & doppler sonography, electrocardiogram (ECG), eye optics, fluids (blood, sweat, tears), imaging sonography, myography, optical coherence tomography, optical computed tomography (CAT) scan, positron emission tomography (PET), radioactivity \& health, ultrasound  & 14 & 19\\  \hline

Microscopy  & atomic force microscopy (AFM), evanescent light scattering, magnetic force microscopy (MFM), scanning electron microscopy (SEM), scanning probe microscopy (SPM), scanning tunneling microscopy (STM), transmission electron microscopy (TEM)  & 13 & 26 \\  \hline

Millikan oil drop  & determining charge of electron  & 7 & 7 \\  \hline

Nuclear magnetic resonance  & electron spin resonance, Larmor precession, nuclear magnetic resonance   & 10 & 14\\  \hline

Optics (advanced)  & acousto-optic modulator (AOM), Berry phase (Pancharatnam phase), critical opalescence, dynamic light scattering, Fabry-Perot cavity, fluorescence correlation spectroscopy, Fourier optics, heterodyning, ion traps, laser interference lithography, magneto-optical trap (MOT), magneto-optic effects, optical fibers (NOT fiber lasers), optical pumping, optical trapping and/or tweezers, photocarrier grating, photoluminescence quantum yield, photomultiplier tube (PMT), photon transfer functions, pump-probe (including femtosecond), quantum cryptography, quantum experiments, Raman-Nath diffraction (acousto-optic diffraction, AOD), single photon correlation, single photon detectors, sonoluminescence, spatial light modulator, wavefront shaping, Zeeman effect  & 17 & 49\\  \hline

Optics (intermediate)  & birefringence, diffraction, greenhouse effect, holography, interferometry and interference (including Young's experiment), microwave diffraction, microwave reflection, microwave scattering, Newton's rings, photometry, physical optics, prisms, refraction (Snell's law), schematic diagrams, spatial filtering, spectroscopy (optical), thin films, wavelengths of visible light  & 68 & 108 \\  \hline

Optics (simple)  & alignment, building a light microscope and/or Kohler's illumination principle, geometric optics, HeNe lasers, lamps, lenses, light sources, mirrors, polarization and/or Brewster's angle, rail optics, ray optics, telescopes and/or galileoscopes   & 23 &39  \\  \hline

Particle physics  & accelerator physics, alpha rays, angular correlation, beta rays, chain reactions, cloud chamber, coincidence measurements, compton scattering, cosmic ray muons,  cross-section, dark matter detection, Fe57 metastable state lifetime, gamma absorption and/or attenuation, gamma spectroscopy,  mass of neutron, Mössbauer effect and/or spectroscopy, muon lifetime, nuclear power, particle tracking, positron emission tomography (PET), relativistic electrons, spectroscopy, strangeness, Z0 decays  & 18 & 44 \\  \hline

Pendulum  & chaotic, coupled, Kater's, physical, Pohl, reversion, simple, torsional  & 23 & 33 \\  \hline

Photoelectric effect  &   & 9 & 9\\  \hline

Plotting  & graphical presentation of measurements, graphing motion, graphing with Excel, graphs, plotting, presenting data   & 5 & 5\\  \hline

Radioactivity  & radioactivity, half-life, attenuation  & 14 & 17 \\  \hline

Solar cells  &   & 6 & 6\\  \hline

Spectroscopy  & atomic spectra, Balmer series, dynamic light scattering, fluorescence correlation spectroscopy, Fourier transform infrared (FTIR) spectroscopy, laser spectroscopy, mass spectroscopy and/or spectrometry, optical grating spectroscopy, Raman spectroscopy, rubidium saturation spectroscopy, saturation spectroscopy, spectroscopy, time-resolved absorption spectroscopy, time-resolved fluorescence spectroscopy; NOT gamma spectroscopy, NOT Mössbauer spectroscopy  & 28 & 37 \\  \hline

Speed of light  &   & 5 & 5\\  \hline

Speed of sound  & measurement of Doppler effect, properties of sound waves, speed of sound in air and in materials  & 8  & 8 \\  \hline

Springs  & Hooke's law, spring constant  & 8 & 9 \\  \hline

Stern-Gerlach experiment &   & 3 & 3 \\  \hline

Test and measurement equipment  & calibration, drift chambers, field programmable gate arrays (FPGAs), image processing, lock-in amplifiers, microcontrollers, micrometers, oscilloscopes, Palmer caliper, periodic signals, reading seismic data, slider caliper, strain gauges, thermocouples/thermometers/thermistors, transducers, Vernier calipers  & 27 & 30\\  \hline

Thermodynamics  &  adiabatic experiments, Boltzmann constant, Boyle's law (Boyle-Mariotte law), calorimetry, critical point, evaporation in a vacuum, heat capacity, heat capacity ratio (Cp/Cv), heat conduction, heat engine, heat pump, heat transfer, heat of combustion, heat of fusion, heat of vaporization (including of liquid nitrogen), ideal gas, linear expansion, Newton's law of cooling, phase transitions, Piston effect, solar cooking box, specific heat, Stirling cycle, temperature dependence of surface tension, thermal expansion, triple points, water vapor  & 26 & 56 \\  \hline

Viscosity  & determination of viscosity of fluid, free fall of sphere in viscous fluid  & 9 & 11\\  \hline

Waves  & electrical waves, mechanical vibrations, properties of sound waves, resonance of electromagnetic waves, sound frequency measurement, sound resonance in open-end tube, standing electromagnetic waves, standing waves, thermal waves, traveling waves, vibrating strings, vibrations, water waves  & 17 &22 \\  \hline 

X-ray experiments  & X-ray diffraction, X-ray experiments  & 12 & 15 \\
        
        \hline \hline

\end{longtable*}

\end{document}


\centering
\textbf{ \LARGE Laboratory Taxonomy Survey}
\vspace{2mm}

\hrule \vspace{1mm} \hrule

\vspace{5mm}

\raggedright

\textbf{Lab Taxonomy Project}
\vspace{2mm}

Thank you for participating in this survey! We are hoping to collect information from all physics \textbf{undergraduate} lab courses around the world. \textbf{Please feel free to send the survey link to any other physics lab course instructors you know.}

In this survey, we will ask you about the course(s) you teach. \textbf{Fill out this survey for ONLY ONE course at a time - if you teach more than one course, fill out the survey once for EACH course you teach.} There is a text box at the end of the survey for additional comments you feel were not captured by the survey questions.

\vspace{2mm}

\hrule

\vspace{5mm}

\textbf{Characteristics of the Institution and Course}

\vspace{2mm}

Institution Name (no abbreviations): \underline{\hspace{2in}}
\vspace{2mm}

What country is the institution located in: \textit{Dropdown List of Countries}
\vspace{2mm}

What is the highest level of degree that is awarded by your institution?
\vspace{-2mm}
   \begin{itemize}[itemsep=-.3ex]
            \item[$\bigcirc$] Associate's degree (2-year college, community college)
            \item[$\bigcirc$] Bachelor's degree
            \item[$\bigcirc$] Master's degree
            \item[$\bigcirc$] PhD
    \end{itemize}
 
\vspace{2mm}

\textbf{Please answer the following questions for ONE course.} If you teach more than one course, please take this survey once per course.

\vspace{2mm}

Course Name (ex. Introduction to Mechanics): \underline{\hspace{2in}}

Course Number, if available (ex. PHYS1140):  \underline{\hspace{2in}}

\vspace{3mm}
What is the level of your course?

Select \textbf{introductory} if your course is part of an introductory physics sequence.

Select \textbf{beyond introductory} if your course is \textbf{NOT} part of the introductory physics sequence \textbf{AND} has at least one prerequisite physics lab course.
\vspace{-2mm}
   \begin{itemize}[itemsep=-.3ex]
            \item[$\bigcirc$] Introductory
            \item[$\bigcirc$] Beyond Introductory
    \end{itemize}

\vspace{2mm}

Approximately how many students are in the entire course in a typical term/semester (all sections)?  \underline{\hspace{1.5in}}

\vspace{2mm}

How many students in total are present in the lab rooms at one time as part of this course (i.e., number of students in a lab section/session)? \underline{\hspace{1.5in}}

\vspace{3mm}

What is the modality of teaching? 
\vspace{-2mm}
   \begin{itemize}[itemsep=-.3ex]
            \item[$\bigcirc$] Mostly/only in person
            \item[$\bigcirc$] Mostly/only online/remote
            \item[$\bigcirc$] Hybrid of in person and online
    \end{itemize}

\vspace{2mm}

What topics in physics are included in the lab portion of the course? \textbf{Select all that apply.}
\vspace{2mm}

 \begin{minipage}{.4\textwidth}
        \begin{itemize}[itemsep=-.1ex]
            \item[$\square$] Astronomy/Astrophysics
            \item[$\square$] Biophysics
            \item[$\square$] Classical Mechanics
            \item[$\square$] Condensed matter/solid state/materials
            \item[$\square$] Electricity and Magnetism
            \item[$\square$] Electronics (analog and/or digital)
            \item[$\square$] Fluid Mechanics
            \item[$\square$] Geophysics
            \item[$\square$] Modern physics
        \end{itemize}
    \end{minipage}
    \begin{minipage}{.4\textwidth}
        \begin{itemize}[itemsep=-.1ex]
            \item[$\square$] Nuclear physics
            \item[$\square$] Optics/ laser physics
            \item[$\square$] Particle physics
            \item[$\square$] Plasma physics
            \item[$\square$] Quantum mechanics
            \item[$\square$] Quantum Information
            \item[$\square$] Thermodynamics
            \item[$\square$] Waves
            \item[$\square$] Other: \underline{\hspace{1.5in}}
        \end{itemize}
    \end{minipage}

\clearpage

\vspace{3mm}

Is the lab course part of a specific physics theory course? 
\vspace{-2mm}
   \begin{itemize}[itemsep=-.3ex]
            \item[$\bigcirc$] Yes
            \item[$\bigcirc$] No
    \end{itemize}

\vspace{2mm}
Does the lab course include lectures on statistics, data analysis, or experimental techniques required for the lab?
\vspace{-2mm}
   \begin{itemize}[itemsep=-.3ex]
            \item[$\bigcirc$] Yes
            \item[$\bigcirc$] No
    \end{itemize}

\vspace{2mm}
Does the lab course meet weekly?
\vspace{-2mm}
   \begin{itemize}[itemsep=-.3ex]
            \item[$\bigcirc$] Yes
            \item[$\bigcirc$] No
    \end{itemize}

\textit{If yes to `does the lab course meet weekly':}
\vspace{1mm}

\hspace{10 mm} How many weeks does the lab course run for each term/semester? \textit{Dropdown 1 - 52}

\vspace{3mm}

\hspace{10 mm} How many hours \textbf{per week} are students scheduled to be in the lab room (working on a lab activity)? \textit{Dropdown 1 - 40}

\vspace{3mm}

\hspace{10 mm} How many hours \textbf{per week} do you estimate students spend in the lab room (working on a lab activity) \textbf{beyond the} 

\hspace{10 mm} \textbf{scheduled time}?

\vspace{2mm}

    \begin{minipage}{.25\textwidth}
        \begin{itemize}[itemsep=-.3ex] \setlength{\itemindent}{1in}
            \item[$\bigcirc$] 0 hours
            \item[$\bigcirc$] 1-3 hours
            \item[$\bigcirc$] 4-6 hours
        \end{itemize}
        \end{minipage}
        \begin{minipage}{.25\textwidth}
            \begin{itemize}[itemsep=-.3ex] 
                \item[$\bigcirc$] More than 6 hours
                \item[$\bigcirc$] Unknown
                \item[$\bigcirc$] Not allowed
            \end{itemize}
        \end{minipage}

\vspace{5mm}

\textit{If no to `does the lab course meet weekly':}

\hspace{10 mm}  Please tell us about the frequency of lab course meetings, number of hours students are scheduled to spend in the lab working 

\hspace{10 mm} on experiments during the course, etc.: \underline{\hspace{3in}}

\vspace{3mm}

Do students have a choice from which experiments they complete?
\vspace{-2mm}
   \begin{itemize}[itemsep=-.3ex]
            \item[$\bigcirc$] Yes, for ALL parts of the course
            \item[$\bigcirc$] Yes, for SOME parts of the course
            \item[$\bigcirc$] No
    \end{itemize}

\vspace{2mm}

Which of the following describes your lab course? A student does... (\textbf{select all that apply})
\vspace{-2mm}
    \begin{itemize}[itemsep=-.1ex]
            \item[$\square$] Multiple experiments per individual lab meeting
            \item[$\square$] One experiment per individual lab meeting
            \item[$\square$] One experiment for multiple individual lab meetings
            \item[$\square$] Multi-session open-ended/partially open-ended project
    \end{itemize}

\vspace{4mm}

\textit{If selected `Multi-session open-ended/partially open-ended project' above:}

\vspace{2mm}

\hspace{10mm} Thinking specifically about the project component of the course...

\vspace{2mm}

\hspace{20mm} How many weeks are spent on project work? \textit{Dropdown 1 - 52}

\vspace{2mm}

\hspace{20mm} Do students choose their own question to investigate?

\vspace{-2mm}
   \begin{itemize}[itemsep=-.3ex]\setlength{\itemindent}{1in}
            \item[$\bigcirc$] Yes
            \item[$\bigcirc$] Sometimes
            \item[$\bigcirc$] No
    \end{itemize}

\vspace{2mm}

\hspace{20mm} Do students design their own experimental procedure?

\vspace{-2mm}
   \begin{itemize}[itemsep=-.3ex]\setlength{\itemindent}{1in}
            \item[$\bigcirc$] Yes
            \item[$\bigcirc$] Sometimes
            \item[$\bigcirc$] No
    \end{itemize}

\vspace{2mm}

\hspace{20mm} Do students build their own experimental setup/apparatus?

\vspace{-2mm}
   \begin{itemize}[itemsep=-.3ex]\setlength{\itemindent}{1in}
            \item[$\bigcirc$] Yes
            \item[$\bigcirc$] Sometimes
            \item[$\bigcirc$] No
    \end{itemize}

\hrule

\clearpage

\vspace{5mm}

\textbf{Students in the Course}

\vspace{2mm}

Students in this course are typically earning a degree in (\textbf{select all that apply}):

\vspace{2mm}

 \begin{minipage}{.4\textwidth}
        \begin{itemize}[itemsep=-.1ex]
            \item[$\square$] Physics
            \item[$\square$] Astrophysics/Astronomy
            \item[$\square$] Physics/Astronomy teaching/pedagogy
            \item[$\square$] Engineering
            \item[$\square$] Mathematics
        \end{itemize}
    \end{minipage}
    \begin{minipage}{.5\textwidth}
        \begin{itemize}[itemsep=-.1ex]
            \item[$\square$] Computer Science
            \item[$\square$] Chemistry
            \item[$\square$] Another science (e.g., biology, geology)
            \item[$\square$] Other teaching/pedagogy
            \item[$\square$] Other degree than above: \underline{\hspace{1.5in}}
        \end{itemize}
    \end{minipage}

\vspace{4mm}

What fraction of students in the course are earning a degree in physics/astrophysics?
\vspace{-2mm}
   \begin{itemize}[itemsep=-.3ex]
            \item[$\bigcirc$] 0-25\%
            \item[$\bigcirc$] 25-50\%
            \item[$\bigcirc$] 50-75\%
            \item[$\bigcirc$] 75-100\%
    \end{itemize}

\vspace{1mm}

Students in the course are typically in their (\textbf{select all that apply - undergraduate only}):
\vspace{-2mm}
   \begin{itemize}[itemsep=-.3ex]
            \item[$\square$] 1st year
            \item[$\square$] 2nd year
            \item[$\square$] 3rd year
            \item[$\square$] 4th year
            \item[$\square$] 5th year or higher
    \end{itemize}

\vspace{2mm}

\hrule

\vspace{4mm}

\textbf{Grouping}

\vspace{3mm}

Do students typically work on lab activities/experiments...
\vspace{-2mm}
   \begin{itemize}[itemsep=-.3ex]
            \item[$\bigcirc$] Alone
            \item[$\bigcirc$] With at least one other student
    \end{itemize}

\vspace{3mm}

\textit{If `With at least one other student' above:}
\vspace{2mm}

\hspace{10mm} How many students are typically in each group:
\vspace{-2mm}
   \begin{itemize}[itemsep=-.3ex]\setlength{\itemindent}{1in}
            \item[$\bigcirc$] 2
            \item[$\bigcirc$] 3
            \item[$\bigcirc$] 4
            \item[$\bigcirc$] 5 or more
    \end{itemize}
\vspace{1mm}

\hspace{10mm} Which of the following best describes the groups in the course?
\vspace{-2mm}
   \begin{itemize}[itemsep=-.3ex]\setlength{\itemindent}{1in}
            \item[$\bigcirc$] Most students stay in the same group for the entire course
            \item[$\bigcirc$] Most students change groups at least once during the term/semester
    \end{itemize}
\vspace{1mm}

\hspace{10mm} Students (\textbf{select all that apply}):
\vspace{-2mm}
   \begin{itemize}[itemsep=-.3ex]\setlength{\itemindent}{1in}
            \item[$\square$] Choose their own group members
            \item[$\square$] Are assigned group members
    \end{itemize}
\vspace{2mm}

\hrule

\vspace{4mm} 
\textbf{Instructional Staff Role}

\vspace{1mm}

Here we are asking about the role of teachers in the lab. These are often referred to using different terms, e.g., instructors, demonstrators, teachers, teaching assistants...

\vspace{3mm}

Number of instructional staff present in the lab at one time:

\hspace{10mm} Faculty/Instructor: \textit{Dropdown 1 - 15+}

\hspace{10mm} Lab Technicians: \textit{Dropdown 1 - 15+}

\hspace{10mm} Postdoc/PhD/Master's teaching assistants: \textit{Dropdown 1 - 15+}

\hspace{10mm} Undergraduate teaching assistants: \textit{Dropdown 1 - 15+}

\clearpage

\textit{If `Postdoc/PhD/Master's teaching assistants' and/or `Undergraduate teaching assistants' are greater than zero above:}

\vspace{2mm}

\hspace{10mm} Training provided to student teaching assistants (\textbf{select all that apply}):
\vspace{-2mm}
   \begin{itemize}[itemsep=-.3ex]\setlength{\itemindent}{1in}
            \item[$\square$] Pedagogy (teaching practices)
            \item[$\square$] Familiarization with lab equipment (e.g., rehearsing the experiment)
            \item[$\square$] How to grade students' work
            \item[$\square$] Other: \underline{\hspace{1.5in}}
    \end{itemize}
\vspace{2mm}

\hspace{10mm} Frequency of training provided to student teaching assistants:
\vspace{-2mm}
   \begin{itemize}[itemsep=-.3ex]\setlength{\itemindent}{1in}
            \item[$\square$] Never
            \item[$\square$] Once per academic year
            \item[$\square$] Once per term/semester
            \item[$\square$] Weekly
            \item[$\square$] Other: \underline{\hspace{1.5in}}
    \end{itemize}
\vspace{2mm}

\hrule

\vspace{4mm} 
\textbf{Goals}

\vspace{3mm}

Which of the following are goals for your lab course?

\begin{table}[H]
    \label{tab:goalsQuestion}
    \begin{tabular}{m{3.3in} | cccc}  \hline\hline 
    \hspace{3mm} & Major Goal & Minor Goal & Not a Goal & Future Goal \\
     & & & & (not currently a goal) \\ \hline \vspace{2mm}
        Reinforcing physics concepts previously seen in lecture (confirming known results / seeing theory in an experiment)  & $\bigcirc$ & $\bigcirc$ & $\bigcirc$ & $\bigcirc$ \\ \vspace{2mm}
       Learning/discovering physics concepts not previously seen in lecture & $\bigcirc$ & $\bigcirc$ & $\bigcirc$ & $\bigcirc$ \\ \vspace{2mm}
        Developing technical knowledge and skills (e.g., making measurements and hands-on manipulation of equipment)  & $\bigcirc$ & $\bigcirc$ & $\bigcirc$ & $\bigcirc$ \\ \vspace{2mm}
        Designing experiments  & $\bigcirc$ & $\bigcirc$ & $\bigcirc$ & $\bigcirc$ \\ \vspace{2mm}
        Develop mathematical model(s) of experimental results  & $\bigcirc$ & $\bigcirc$ & $\bigcirc$ & $\bigcirc$ \\ \vspace{2mm}
        Learning how to analyze and interpret data (e.g., linear regressions, uncertainty)  & $\bigcirc$ & $\bigcirc$ & $\bigcirc$ & $\bigcirc$ \\ \vspace{2mm}
        Learning how to visualize data (e.g., plotting) & $\bigcirc$ & $\bigcirc$ & $\bigcirc$ & $\bigcirc$ \\ \vspace{2mm}
        Developing lab notebook keeping skills & $\bigcirc$ & $\bigcirc$ & $\bigcirc$ & $\bigcirc$ \\
        Developing scientific writing skills (e.g., lab reports)  & $\bigcirc$ & $\bigcirc$ & $\bigcirc$ & $\bigcirc$ \\ \vspace{2mm}
        Developing other communication skills  (e.g., oral presentations, poster presentations)
        & $\bigcirc$ & $\bigcirc$ & $\bigcirc$  & $\bigcirc$ \\ \vspace{2mm}
        Making quick and simple approximations to predict experimental outcomes (e.g., back of the envelope calculations)  & $\bigcirc$ & $\bigcirc$ & $\bigcirc$  & $\bigcirc$ \\ \vspace{2mm}
        Developing expert-like views about the nature of the process of doing experimental physics (e.g., experimentation is iterative, not linear) & $\bigcirc$ & $\bigcirc$ & $\bigcirc$ & $\bigcirc$  \\ \vspace{2mm}
        Developing collaboration and teamwork skills & $\bigcirc$ & $\bigcirc$ & $\bigcirc$ & $\bigcirc$ \\ \vspace{2mm}
        Reflecting on and evaluating one’s own learning  and knowledge (metacognition) & $\bigcirc$ & $\bigcirc$ & $\bigcirc$ & $\bigcirc$ \\  \vspace{2mm}
        Enjoying experimental physics and/or the course & $\bigcirc$ & $\bigcirc$ & $\bigcirc$ & $\bigcirc$ \\ \\  \hline\hline
    \end{tabular}
\end{table}

\vspace{3mm}
\hrule

\clearpage 

\vspace{4mm} 
\textbf{Activities}

\vspace{3mm}

Which, if any, of the following \textbf{officially branded} approaches to lab instruction do you use in whole or part in your class? \textbf{Select all that apply.}

If you don't recognize these, please select ``None of the above''.

Feel free to click the links to learn more about each type of instruction. \textit{Note: in the actual survey, each of these options is a link to a page with more information about each of these types of instruction}

\vspace{-2mm}
   \begin{itemize}[itemsep=-.3ex]\setlength{\itemindent}{1in}
            \item[$\square$] ISLE Physics
            \item[$\square$] SCALE-UP
            \item[$\square$] Modeling Instruction
            \item[$\square$] Studio Physics
            \item[$\square$] Course-based undergraduate research experience (CURE)
            \item[$\square$] Thinking Critically in Physics Labs (Cornell Designed Labs)
            \item[$\square$] Other: \underline{\hspace{1.5in}}
            \item[$\bigcirc$] None of the above
    \end{itemize}
\vspace{2mm}

Do you often use any of the following research-based assessments to evaluate teh course? \textbf{Select all that apply.}

Feel free to click the links to learn more about each assessment. \textit{Note: in the actual survey, each of these options is a link to a page with more information about each assessment.}

\vspace{-2mm}
   \begin{itemize}[itemsep=-.3ex]\setlength{\itemindent}{1in}
            \item[$\square$] Colorado Learning Attitudes about Science Survey for Experimental Physics (E-CLASS)
            \item[$\square$] Modeling Assessment for Physics Laboratory Experiments (MAPLE)
            \item[$\square$] Physics Lab Inventory of Critical thinking (PLIC)
            \item[$\square$] Survey of Physics Reasoning on Uncertainty Concepts in Experiments (SPRUCE)
            \item[$\square$] Other: \underline{\hspace{1.5in}}
            \item[$\bigcirc$] None of the above
    \end{itemize}
\vspace{5mm}

On average for the course, how often do students engage with the following activities?

\vspace{3mm}

\textbf{Data Analysis and Visualization}

\begin{table}[H]
    \label{tab:dataAnalysisQuestion}
    \begin{tabular}{m{3.3in} | ccccc}  \hline\hline 
    \hspace{3mm} & Never & Would like to use & 1-2  times per & Somewhat & Very \\
     & & in the future &semester/term & frequently & frequently  \\ \hline \vspace{2mm}
        Quantify uncertainty in a measurement  & $\bigcirc$ & $\bigcirc$ & $\bigcirc$ & $\bigcirc$ & $\bigcirc$ \\ \vspace{2mm}
       Calculate uncertainty using error propagation & $\bigcirc$ & $\bigcirc$ & $\bigcirc$ & $\bigcirc$  & $\bigcirc$  \\ \vspace{2mm}
       Use software to aid with data analysis and visualization & $\bigcirc$ & $\bigcirc$ & $\bigcirc$ & $\bigcirc$  & $\bigcirc$  \\ \vspace{2mm} 
        Fit a mathematical model to data (e.g., linear regression) & $\bigcirc$ & $\bigcirc$ & $\bigcirc$ & $\bigcirc$  & $\bigcirc$  \\ \vspace{2mm}
        Relate curve fitting parameters to physical quantities & $\bigcirc$ & $\bigcirc$ & $\bigcirc$ & $\bigcirc$  & $\bigcirc$  \\ \vspace{2mm}
        Write their own code to analyze data (e.g., in Python, Mathematica, Matlab) & $\bigcirc$ & $\bigcirc$ & $\bigcirc$ & $\bigcirc$  & $\bigcirc$  \\ \vspace{2mm}
        Determine the mathematical model to best represent the data  & $\bigcirc$ & $\bigcirc$ & $\bigcirc$ & $\bigcirc$  & $\bigcirc$  \\ \vspace{2mm}
        Determine if two measurements (with uncertainty) are consistent with each other & $\bigcirc$ & $\bigcirc$ & $\bigcirc$ & $\bigcirc$  & $\bigcirc$  \\ 
         \hline\hline
    \end{tabular}
\end{table}

\clearpage

\vspace{3mm}

\textbf{Communication}

\begin{table}[H]
    \label{tab:communicationQuestion}
    \begin{tabular}{m{3.3in} | ccccc}  \hline\hline 
    \hspace{3mm} & Never & Would like to use & 1-2  times per & Somewhat & Very \\
     & & in the future &semester/term & frequently & frequently  \\ \hline \vspace{2mm}
        Give oral presentations as an individual  & $\bigcirc$ & $\bigcirc$ & $\bigcirc$ & $\bigcirc$ & $\bigcirc$ \\ \vspace{2mm}
       Give oral presentations as a group & $\bigcirc$ & $\bigcirc$ & $\bigcirc$ & $\bigcirc$  & $\bigcirc$  \\ \vspace{2mm}
       Write lab reports & $\bigcirc$ & $\bigcirc$ & $\bigcirc$ & $\bigcirc$  & $\bigcirc$  \\ \vspace{2mm} 
        Design and present a poster & $\bigcirc$ & $\bigcirc$ & $\bigcirc$ & $\bigcirc$  & $\bigcirc$  \\ \vspace{2mm}
        Maintain an individual lab notebook (paper or electronic) & $\bigcirc$ & $\bigcirc$ & $\bigcirc$ & $\bigcirc$  & $\bigcirc$  \\ \vspace{2mm}
        Maintain a group lab notebook (paper or electronic) & $\bigcirc$ & $\bigcirc$ & $\bigcirc$ & $\bigcirc$  & $\bigcirc$  \\ \vspace{2mm}
        Do a literature search  & $\bigcirc$ & $\bigcirc$ & $\bigcirc$ & $\bigcirc$  & $\bigcirc$  \\ \vspace{2mm}
        Read scientific papers & $\bigcirc$ & $\bigcirc$ & $\bigcirc$ & $\bigcirc$  & $\bigcirc$  \\ 
        Write a proposal for designing or conducting an experiment  & $\bigcirc$ & $\bigcirc$ & $\bigcirc$ & $\bigcirc$ & $\bigcirc$ \\ \vspace{2mm}
        Complete a worksheet/fill-in-the-blank template  & $\bigcirc$ & $\bigcirc$ & $\bigcirc$ & $\bigcirc$ & $\bigcirc$ \\ \vspace{2mm}
        Provide peer feedback to other students in the course  & $\bigcirc$ & $\bigcirc$ & $\bigcirc$ & $\bigcirc$ & $\bigcirc$ \\ 
         \hline\hline
    \end{tabular}
\end{table}

\vspace{3mm}

\textbf{Student Decision-Making}

\begin{table}[H]
    \label{tab:agencyQuestion}
    \begin{tabular}{m{3.3in} | ccccc}  \hline\hline 
    \hspace{3mm} & Never & Would like to use & 1-2  times per & Somewhat & Very \\
     & & in the future &semester/term & frequently & frequently  \\ \hline \vspace{2mm}
        Develop their own research questions  & $\bigcirc$ & $\bigcirc$ & $\bigcirc$ & $\bigcirc$ & $\bigcirc$ \\ \vspace{2mm}
       Design their own procedures for data collection & $\bigcirc$ & $\bigcirc$ & $\bigcirc$ & $\bigcirc$  & $\bigcirc$  \\ \vspace{2mm}
      Design their own apparatus& $\bigcirc$ & $\bigcirc$ & $\bigcirc$ & $\bigcirc$  & $\bigcirc$  \\ \vspace{2mm} 
        Choose their own analysis methods & $\bigcirc$ & $\bigcirc$ & $\bigcirc$ & $\bigcirc$  & $\bigcirc$  \\ \vspace{2mm}
        Refine experimental apparatus or procedure to reduce uncertainty (statistical and/or systematic) & $\bigcirc$ & $\bigcirc$ & $\bigcirc$ & $\bigcirc$  & $\bigcirc$  \\ 
         \hline\hline
    \end{tabular}
\end{table}

\vspace{3mm}

\textbf{Materials and Resources}

\begin{table}[H]
    \label{tab:materialsQuestion}
    \begin{tabular}{m{3.3in} | ccccc}  \hline\hline 
    \hspace{3mm} & Never & Would like to use & 1-2  times per & Somewhat & Very \\
     & & in the future &semester/term & frequently & frequently  \\ \hline \vspace{2mm}
        Build their own apparatus using equipment provided  & $\bigcirc$ & $\bigcirc$ & $\bigcirc$ & $\bigcirc$ & $\bigcirc$ \\ \vspace{2mm}
       Build their own apparatus, using equipment they have at home & $\bigcirc$ & $\bigcirc$ & $\bigcirc$ & $\bigcirc$  & $\bigcirc$  \\ \vspace{2mm}
      Use pre-constructed apparatus & $\bigcirc$ & $\bigcirc$ & $\bigcirc$ & $\bigcirc$  & $\bigcirc$  \\ \vspace{2mm} 
        Use a step-by-step lab manual to complete experiments & $\bigcirc$ & $\bigcirc$ & $\bigcirc$ & $\bigcirc$  & $\bigcirc$  \\ \vspace{2mm}
        Use a semi-guided (less guidance than step-by-step) lab manual to complete experiments & $\bigcirc$ & $\bigcirc$ & $\bigcirc$ & $\bigcirc$  & $\bigcirc$  \\  \vspace{2mm}
        Use a scientific paper to guide lab experiments & $\bigcirc$ & $\bigcirc$ & $\bigcirc$ & $\bigcirc$  & $\bigcirc$  \\  \vspace{2mm}
        Write code to collect data during experiments (e.g., LabVIEW/Python) & $\bigcirc$ & $\bigcirc$ & $\bigcirc$ & $\bigcirc$  & $\bigcirc$  \\  \vspace{2mm}
        Use commercial educational equipment (e.g., Vernier, Pasco, Teachspin) & $\bigcirc$ & $\bigcirc$ & $\bigcirc$ & $\bigcirc$  & $\bigcirc$  \\  
                        
         \hline\hline
    \end{tabular}
\end{table}

\clearpage

\vspace{3mm}

\textbf{Modeling and Other Activities}

\begin{table}[H]
    \label{tab:materialsQuestion}
    \begin{tabular}{m{3.3in} | ccccc}  \hline\hline 
    \hspace{3mm} & Never & Would like to use & 1-2  times per & Somewhat & Very \\
     & & in the future &semester/term & frequently & frequently  \\ \hline \vspace{2mm}
        Use mathematical or conceptual models to make predictions  & $\bigcirc$ & $\bigcirc$ & $\bigcirc$ & $\bigcirc$ & $\bigcirc$ \\ \vspace{2mm}
       Calibrate measurement tools & $\bigcirc$ & $\bigcirc$ & $\bigcirc$ & $\bigcirc$  & $\bigcirc$  \\ \vspace{2mm}
      Confirm results previously known to students & $\bigcirc$ & $\bigcirc$ & $\bigcirc$ & $\bigcirc$  & $\bigcirc$  \\ \vspace{2mm} 
        Determine results already known to instructors but not known by students before the lab & $\bigcirc$ & $\bigcirc$ & $\bigcirc$ & $\bigcirc$  & $\bigcirc$  \\ \vspace{2mm}
        Determine results unknown to both students and instructors before the labs (e.g., coefficient of friction of unknown surfaces) & $\bigcirc$ & $\bigcirc$ & $\bigcirc$ & $\bigcirc$  & $\bigcirc$  \\  \vspace{2mm}
        Troubleshoot problems with the setup or apparatus & $\bigcirc$ & $\bigcirc$ & $\bigcirc$ & $\bigcirc$  & $\bigcirc$  \\  \vspace{2mm}
        Watch a video (e.g., LabVIEW/Python) & $\bigcirc$ & $\bigcirc$ & $\bigcirc$ & $\bigcirc$  & $\bigcirc$  \\  \vspace{2mm}
        Complete safety training  & $\bigcirc$ & $\bigcirc$ & $\bigcirc$ & $\bigcirc$  & $\bigcirc$  \\ \vspace{2mm}  
        Engage with PhET simulations   & $\bigcirc$ & $\bigcirc$ & $\bigcirc$ & $\bigcirc$  & $\bigcirc$  \\
                        
         \hline\hline
    \end{tabular}
\end{table}

\hrule

\vspace{5mm} 
\textbf{Evaluating Student Learning}

\vspace{3mm}

\textit{If `With at least one other student' is selected in the question about whether students work alone or with others:}

\vspace{3mm}

\hspace{10mm} Are students graded/evaluated individually?

\vspace{-2mm}
   \begin{itemize}[itemsep=-.3ex]\setlength{\itemindent}{1in}
            \item[$\bigcirc$] Yes
            \item[$\bigcirc$] Yes, but some part of the grade is the same for all members of a group
            \item[$\bigcirc$] No, they are given a grade only for their group
    \end{itemize}

\vspace{3mm}

Select all of the following that are marked/graded to determine the final course grade (\textbf{select all the apply}):

\vspace{2mm}

 \begin{minipage}{.53\textwidth}
        \begin{itemize}[itemsep=-.1ex]
            \item[$\square$] Before the lab (prelab) calculations, questions, and/or worksheet
            \item[$\square$] Taking a quiz at home before the lab
            \item[$\square$] Writing a measurement and/or analysis plan before the lab
            \item[$\square$] Watching a video before the lab
            \item[$\square$] Quiz/interview in the lab prior to being allowed to work
            \item[$\square$] Attendance and/or participation
            \item[$\square$] Lab notebooks
            \item[$\square$] Worksheets/fill-in-the-blank template completed during the lab
            \item[$\square$] Lab report
            \item[$\square$] Partial lab report (e.g., methods section only, results section only)
        \end{itemize}
    \end{minipage}
    \begin{minipage}{.45\textwidth}
        \begin{itemize}[itemsep=-.1ex]
            \item[$\square$] Accuracy and/or precision of experimental results
            \item[$\square$] Oral presentation
            \item[$\square$] Poster presentation
            \item[$\square$] Written exam
            \item[$\square$] Practical exam (i.e., hands-on exam)
            \item[$\square$] Peer feedback on other students' work
            \item[$\square$] Observation of students (e.g., use of equipment or teamwork)
            \item[$\square$] Interview/meeting - discussion of lab between student and instructor
            \item[$\square$] Other: \underline{\hspace{2in}}
        \end{itemize}
    \end{minipage}

\vspace{3mm}

When grading students, do you use a rubric (a set of guidelines about how something is graded)?
\vspace{-2mm}
   \begin{itemize}[itemsep=-.3ex]\setlength{\itemindent}{1in}
            \item[$\bigcirc$] Yes, for all parts that are graded
            \item[$\bigcirc$] Yes, for some parts that are graded
            \item[$\bigcirc$] No
    \end{itemize}

\vspace{3mm}
\hrule

\vspace{3mm}
 
\textbf{Optional Long Entry}

\vspace{3mm}

(Optional) Enter the titles of your lab experiments (use any language): \underline{\hspace{2in}}

\vspace{3mm}

(Optional) Additional comments about your lab course that might not have been captured above: \underline{\hspace{1.75in}}